\documentclass[12pt]{article}
\usepackage{amssymb}
\usepackage{amsmath}
\usepackage{amstext}
\usepackage{graphicx,epsfig}
\usepackage{epsfig}
\usepackage{verbatim} 
\usepackage{fancyhdr}
\usepackage{fancybox}
\usepackage{color}
\usepackage{ulem,bbold}
\usepackage{enumitem}
\usepackage{subfigure}
\usepackage{bbm}
\usepackage{parskip}
\usepackage{pifont}
\usepackage{caption}
\usepackage{adjustbox}

\newcommand{\Comment}[1]{{}}
\definecolor{MyDarkBlue}{rgb}{0.15,0.15,0.45}
\newcommand{\be}{\begin{equation}}  
\newcommand{\ee}{\end{equation}}  
\newcommand{\bea}{\begin{eqnarray}}  
\newcommand{\eea}{\end{eqnarray}}
\newcommand{\nn}{\nonumber}

\newcommand{\hatM}{\hat M}

\def\({\left(}
\def\){\right)}

\def\mpl{M_{\rm Pl}}
\def\p{\partial}

\def\lsim{\mathrel{\rlap{\lower3pt\hbox{\hskip0pt$\sim$}}
     \raise1pt\hbox{$<$}}}         %less than or approx. symbol
\def\gsim{\mathrel{\rlap{\lower4pt\hbox{\hskip1pt$\sim$}}
     \raise1pt\hbox{$>$}}}         %greater than or approx. symbol
\def\lsim{\mathrel{\rlap{\lower3pt\hbox{\hskip0pt$\sim$}}
     \raise1pt\hbox{$<$}}}         %less than or approx. symbol
\def\gsim{\mathrel{\rlap{\lower4pt\hbox{\hskip1pt$\sim$}}
     \raise1pt\hbox{$>$}}}         %greater than or approx. symbol
\def\beq{\begin{eqnarray}}
\def\eeq{\end{eqnarray}}
\def\ba{\begin{eqnarray}}
\def\ea{\end{eqnarray}}
\def\({\left(}
\def\){\right)}

\def\be{\begin{equation}}
\def\ee{\end{equation}}

\def\beq{\begin{equation}}
\def\eeq{\end{equation}}

\newcommand{\x}{\vec{x}}

\def\({\left(}
\def\){\right)}

\def\mpl{M_{\rm Pl}}
\def\p{\partial}

\def\x{\mathbf{x}}

\def\lsim{\mathrel{\rlap{\lower3pt\hbox{\hskip0pt$\sim$}}
     \raise1pt\hbox{$<$}}}         %less than or approx. symbol
\def\gsim{\mathrel{\rlap{\lower4pt\hbox{\hskip1pt$\sim$}}
     \raise1pt\hbox{$>$}}}         %greater than or approx. symbol
\def\lsim{\mathrel{\rlap{\lower3pt\hbox{\hskip0pt$\sim$}}
     \raise1pt\hbox{$<$}}}         %less than or approx. symbol
\def\gsim{\mathrel{\rlap{\lower4pt\hbox{\hskip1pt$\sim$}}
     \raise1pt\hbox{$>$}}}         %greater than or approx. symbol
\def\beq{\begin{eqnarray}}
\def\eeq{\end{eqnarray}}
\def\ba{\begin{eqnarray}}
\def\ea{\end{eqnarray}}
\def\({\left(}
\def\){\right)}

\numberwithin{equation}{section}

\begin{document}

%\maketitle

%\begin{flushright}
%{UCSD-PTH-12-11}
%\end{flushright}

\begin{center}
{\Large\bf {Weakly Broken Galileon Symmetry}}
\vspace{0.5cm}
%\\
%\vspace{0.15cm}
%{\LARGE  { and Effective Field Theory near de Sitter Space }}
%\\
%\vspace{0.15cm}
%{\LARGE  { in Cosmology }}
\end{center} 
 \vspace{0.5truecm}
\thispagestyle{empty} 
\centerline{
David Pirtskhalava${}^{a,}$\footnote{E-mail address: david.pirtskhalava@sns.it}, 
Luca Santoni${}^{a,b,}$\footnote{E-mail address: luca.santoni@sns.it},
Enrico Trincherini${}^{a,b,}$\footnote{E-mail address: enrico.trincherini@sns.it},
Filippo Vernizzi${}^{c,}$\footnote{E-mail address: filippo.vernizzi@cea.fr}
}

\vspace{0.5 cm}

\centerline{\it $^a$Scuola Normale Superiore, Piazza dei Cavalieri 7, 56126, Pisa, Italy}

 \vspace{.1cm}

\centerline{\it $^b$INFN - Sezione di Pisa, 56100 Pisa, Italy}

 \vspace{.1cm}

\centerline{\it $^c$Institut de physique th\' eorique, Universit\'e  Paris Saclay,}
\centerline{\it CEA, CNRS, 91191 Gif-sur-Yvette, France}

\vspace{2cm}
\begin{abstract}

Effective theories of a scalar $\phi$ invariant under the internal \textit{galileon symmetry} $\phi\to\phi+b_\mu x^\mu$ have been extensively studied due to their special theoretical and phenomenological properties.
In this paper, we introduce the notion of \textit{weakly broken galileon invariance}, which characterizes the unique class of couplings of such theories to gravity that maximally retain their defining symmetry. The curved-space remnant of the galileon's quantum properties allows to construct (quasi) de Sitter backgrounds largely insensitive to loop corrections. We exploit this fact to build novel cosmological models with interesting phenomenology, relevant for both inflation and late-time acceleration of the universe.
\end{abstract}

\newpage

\thispagestyle{empty}
%\tableofcontents
\newpage
\setcounter{page}{1}
\setcounter{footnote}{0}

\section{Introduction and summary}

Over the years, the effective field theory (EFT) approach has proven itself invaluable  for organizing physics at different length scales. With the sole knowledge of the relevant physical degrees of freedom and symmetries -- exact or approximate -- that govern their dynamics, EFTs can be used to infer model-independent physical predictions. The reason for such universality lies in a fundamental property of the great majority of physical systems, according to which the details of physics at short distances do not leave a qualitative impact on their large-distance characteristics, enabling one to analyze theories with various UV structure in a single framework. 

In recent years, the EFT approach has offered new insights in various cosmological contexts, including -- but not limited to -- inflation \cite{Creminelli:2006xe,Cheung:2007st}, dark energy \cite{Creminelli:2008wc,Gubitosi:2012hu,Bloomfield:2012ff} and the large-scale structure \cite{Carrasco:2012cv}.   
For example, the effective theory of inflation is based on the observation that the dynamics of the most general theory of `single-clock' inflation can be universally captured by an EFT nonlinearly realizing time diffeomorphisms  (diffs)\footnote{In what follows, we will at times abuse the nomenclature by referring to this redundancy as `time translations'.} $t\to t+\xi^0(t,\x)~,$ with spatial diffs $x^i\to x^i+\xi^i (t,\x)$ realized linearly. The energy scales of interest are those around the inflationary Hubble rate, $E\sim H$, i.e.~the frequency at which all correlation functions are measured in the CMB. The spectrum of perturbations consists of the two polarizations of the graviton plus the Goldstone boson of time translation symmetry breaking. That the latter mode has to be present in the spectrum is a direct consequence of the symmetry breaking pattern and  has little to do with the exact UV details of the microscopic theory of inflation. This Goldstone boson is what we usually refer to as the adiabatic mode. The CMB indicates that precisely this mode is predominantly responsible for the generation of large-scale structures in the universe. Indeed, any UV theory of inflation that does not lead to extra degrees of freedom around the Hubble energies is equivalent to single-clock inflation and is thus captured by the EFT of Refs. \cite{Creminelli:2006xe,Cheung:2007st}.

Given that symmetries define the effective theories, it is of interest to explore the possible symmetries of systems consisting of one or more scalar fields coupled to gravity -- a typical setup for cosmological model-building. For example, in the context of inflation, effective theories with (approximate) shift symmetry play a key role. 
In this paper, we study yet another possible -- and, as we argue below, necessarily approximate -- symmetry of cosmological scalar fields: the invariance under \textit{internal} galileon transformations 
\beq
\label{g_inv}
\phi \to \phi + b_\mu x^\mu~.
\eeq
Theories invariant under \eqref{g_inv} have appeared in various contexts before. To start with, Eq.~\eqref{g_inv} is a symmetry (up to a total derivative) of the simplest possible quantum field theory: that of a free scalar field. The simplest \textit{interacting} generalization, nontrivially invariant under \eqref{g_inv}, i.e.~containing less than two derivatives per field in the Lagrangian, has appeared in Ref.~\cite{Luty:2003vm} in the context of the DGP model \cite{Dvali:2000hr}. The most general scalar theory with the same property -- the \textit{galileon} -- has been proposed in \cite{Nicolis:2008in} and subsequently found \cite{deRham:2010ik} to describe the scalar polarization of the ghost-free dRGT massive graviton \cite{deRham:2010kj}. Galileon theories are interesting in many ways. Their defining property is that, despite being higher-derivative theories, they have no more than second-order equations of motion, thus describing a single propagating scalar mode. Moreover, as a direct consequence of invariance under \eqref{g_inv}, the coefficients of the leading galileon interactions are not renormalized, at least perturbatively \cite{Luty:2003vm}, making any tuning of them stable under quantum corrections. 

Despite these interesting properties, the invariance under galileon transformations cannot be exact in nature because the couplings of the scalar to gravity, at least, break it explicitly. We will therefore be interested in characterizing the theories that preserve as much as possible the attractive quantum non-renormalization properties of the exactly invariant case. This naturally leads to the notion of \textit{weakly broken} invariance under \eqref{g_inv}, which we will define in a precise way in what follows. 
The simplest sufficiently non-trivial theory with \textit{weakly broken galileon invariance} (and exact shift symmetry) is of the following form
\beq
\label{simpleaction}
\mathcal{L}=-\frac{1}{2}(\p\phi)^2 +\frac{1 }{\Lambda_3^3}(\p\phi)^2\Box\phi+\frac{1}{\Lambda_2^4}(\p\phi)^4~.
\eeq
While the first two operators in \eqref{simpleaction} are exactly invariant under \eqref{g_inv} (up to the boundary terms), the quartic operator is a small breaking as far as $\Lambda_2 \gg \Lambda_3$. 
In general we would expect other symmetry breaking operators of the form $(\partial \phi)^{2n}$ to be generated by quantum corrections at the scale $\Lambda_2$. However, 
in this case, a stronger result, the remnant of the non-renormalization properties of the invariant action, holds: all the symmetry breaking operators are generated at a scale that is parametrically higher than $\Lambda_2$. This means in particular that the operator $(\partial \phi)^4$ gets only small corrections through loop effects. In the presence of gravity, as we will show in the next section, the weak breaking gives rise to even more non-trivial consequences.    

Importantly, the theory \eqref{simpleaction} admits a  large generalization that retains both its quantum properties and the second-order equations of motion. Perhaps not surprisingly, we will find that the proper generalization fits into the class of the most general Lorentz-invariant theories with second-order field equations, known as Horndeski theories \cite{Horndeski:1974wa}. Since, however, a generic Horndeski theory does not have much to do with the invariance under \eqref{g_inv}, we prefer to refer to the subclass that we study here as `theories with weakly broken galileon (WBG) invariance'. The purpose of this work is to introduce these theories and to set the stage for their detailed phenomenological studies.

The study of galileon-invariant theories in the inflationary context have been initiated in Ref.~\cite{Burrage:2010cu}. The model, referred to as `galileon inflation', is based on the particular curved-space extension of the theory that keeps the property of second-order equations of motion intact -- the so-called `covariant galileon' of Ref.~\cite{Deffayet:2009wt}. It has been pointed out in \cite{Burrage:2010cu}, that such theories enjoy more freedom compared to ghost/DBI inflation-like models as far as phenomenology is concerned. In particular, they can lead to the possibility of lifting the `large speed of sound/small non-gaussianities' correspondence, characteristic of the latter models. 
Our results extend the findings of Ref.~\cite{Burrage:2010cu} in several directions. While it has been assumed in \cite{Burrage:2010cu} that the covariant galileon is the unique model capable of extending the phenomenological virtues of shift-symmetric theories in a radiatively stable way, we find that there is in fact a wider class of models that can achieve this. Unlike the covariant galileon, the theories we will be interested in do not generically reduce to the standard galileon once gravity is turned off. Nevertheless, neither  non-renormalization nor the second-order equations of motion of the galileon need to be given up, leading to the possibility of strong -- and quantum-mechanically robust -- phenomenological differences from slow-roll inflation. This generalizes the well-studied case of DBI models in an interesting way. 

The paper is organized as follows. We start in Secs.~\ref{sectwo} and \ref{secthree} by showing that a (shift-symmetric) subset of Horndeski theories can be \textit{derived} solely based  on the requirement of WBG invariance at the quantum level, a concept that we define along the way. In Sec.~\ref{secfour}, we study general cosmological implications of approximate symmetry under galileon transformations and apply our findings to inflationary, as well as late-time cosmologies in Secs.~\ref{secfive} and \ref{secsix}, respectively. Finally, in Sec.~\ref{secseven} we conclude. Various technical calculations that would overwhelm the main body of the text are collected in the two appendices.

\section{Galileon invariance (with and without gravity)}
\label{sectwo}
In this section we study the fate of galileon symmetry upon inclusion of gravity, starting out by reviewing flat-space theories \textit{exactly} invariant under \eqref{g_inv}. If internal galileon invariance is realized in nature, however, it has to be approximate: the coupling to gravity unavoidably breaks this symmetry and, at the same time, defines the sense in which the breaking can be considered weak. 
Indeed, as we will see below, out of the infinite number of possible ways one can couple the galileon to gravity, there is a unique set of non-minimal couplings that can be qualified as being `more invariant' under galileon transformations than all the rest. The resulting theory is the \textit{covariant galileon} of Ref.~\cite{Deffayet:2009wt}. On the other hand, since the galileon symmetry has to be weakly broken by the couplings to gravity, one should in principle include in the effective theory symmetry-breaking operators of e.g.~the form $(\p\phi)^{2n}$. The latter will be generated by quantum loops with Wilson coefficients bounded from above, so as to be consistent with the approximate symmetries at hand. As we will see, requiring  invariance under \eqref{g_inv} to be only \textit{weakly} broken will naturally lead to a particular sub-class of Horndeski theories, which is however much more general than just the covariant galileon. 
 
\subsection{Flat space galileons}

Consider a trivial, free theory of a scalar $\phi$. As emphasized above, in addition to more familiar symmetries (such as the ones under constant shifts or conformal transformations), this theory possesses an extra invariance under internal galileon transformations \eqref{g_inv}. The latter leaves the action invariant only up to a boundary term. As shown in \cite{Nicolis:2008in}, apart from the free theory -- and even a more trivial tadpole term, there are exactly three more interaction terms in four spacetime dimensions that share invariance under internal galileon transformations in a non-trivial way. The corresponding theory can be written as 
\beq
\label{gallag}
\mathcal{L}=(\p\phi)^2 +\sum_{I=3}^5\frac{c_a}{\Lambda_3^{3(I-2)}}\mathcal{L}_i ~,
\eeq
with the three interaction terms given by 
\begin{align}
\label{gal1}
\mathcal{L}_3 &= (\p\phi)^2~[\Phi]~, \\
\label{gal2}
\mathcal{L}_4 &= (\p\phi)^2~ \big([\Phi]^2-
[\Phi^2]\big )~,\\
\label{gal3}
\mathcal{L}_5 &= (\p\phi)^2~\big([\Phi]^3-3[\Phi][\Phi^2]+ 2[\Phi^3]\big )~,
\end{align}
where we denote $[\Phi]\equiv \Box\phi,~[\Phi^2]\equiv \p^\mu\p_\nu\phi\p^\nu\p_\mu\phi$, etc.
In addition to being invariant under \eqref{g_inv}, galileon theories share another special property: the associated scalar equations of motion are second order, both in time and in space, despite the higher-derivative interactions in Eqs.~\eqref{gal1}--\eqref{gal3}. This guarantees that there are no extra propagating degrees of freedom hidden in $\phi$. Moreover, the particular structure of the Lagrangian \eqref{gallag} results in a non-renormalization theorem that  allows to radiatively generate only terms trivially invariant under \eqref{g_inv}, i.e.~terms with at least two derivatives acting on $\phi$. In particular, the coefficients $c_I$ of galileon interactions are not renormalized by quantum loops in the presence of exact galileon invariance \cite{Luty:2003vm}. In theories of modified gravity with matter, galileons couple to the rest of the degrees of freedom via symmetry-breaking Planck-suppressed operators\footnote{In massive gravity, where $\phi$ describes the helicity-0 polarization of the graviton, this simply follows from the equivalence principle.}, usually making the renormalization of $c_I$ parametrically weak.

\subsection{Coupling to gravity}

It is generally impossible to couple the galileon to gravity\footnote{Unless gravity is massive, see \cite{Gabadadze:2012tr}.}, while keeping invariance under galileon transformations, or their curved-space generalization, intact. Thus, galileon symmetry is expected to be broken even in the purely scalar sector -- i.e.~if one sets the metric perturbation to zero `by hand' -- due to  loop-generated operators not invariant under \eqref{g_inv} (these operators of course have to be suppressed by at least one power of the Planck mass). For the sake of concreteness, let us concentrate on the operators of the form $(\p\phi)^{2n}$. After all, it is the absence, or smallness, of these operators that makes the galileon special compared to a generic shift-symmetric theory\footnote{Moreover, the size of these operators is an avatar for technical naturalness of non-trivial de Sitter vacua of the classical theory, as we will show in the next section.}. These operators can be generated from loops with the appropriate number of insertions of symmetry-breaking vertices, that include at least one graviton line. 
Interpreting $\Lambda_3$ -- the smallest scale by which interactions are suppressed in the effective theory\footnote{We always assume in this paper that $\Lambda_3\ll\mpl$.} -- as the genuine cutoff, any loop-generated operator can be written as
\beq
\label{oneder}
\frac{(\p\phi)^{2n}}{\Lambda_{k,n}^{4(n-1)}}~, \qquad \Lambda_{k,n}\equiv \left[ \mpl^k \Lambda_3^{4n-k-4}\right]^{\frac{1}{4n-4}}~,
\eeq
where $k$ denotes some positive integer and we have omitted  factors of $16\pi^2$ for simplicity. For  fixed $k$ and $n$ sufficiently large, the scale suppressing a given single-derivative operator, $\Lambda_{k,n} $, can in principle be arbitrarily close to $\Lambda_3$. If this were true, the resulting theory would by no means be considered as a theory with WBG invariance, since symmetry-breaking operators would be order-one in the units of the cutoff.

We wish to show here that: i) this in fact does not happen for the galileon, even if minimally coupled to gravity; ii) for a very particular, non-minimal coupling, the scale suppressing the symmetry-breaking operators can be made parametrically higher than the analogous scale characterizing all the other ways of introducing gravity into the system. For the latter theories, there is a well-defined separation between the scale suppressing the invariant galileon interactions and the quantum-mechanically generated single-derivative operators: while the former are suppressed by powers of $\Lambda_3$, the scale suppressing the latter is parametrically higher, asymptoting to 
\beq
\Lambda_2\equiv (\mpl \Lambda_3^3)^{1/4} 
\eeq
for large $n$ in \eqref{oneder}. Perhaps not surprisingly, we will find that the special type of couplings to gravity required to suppress symmetry-breaking operators is of the Horndeski class -- the unique curved-space extension of the (generalized) galileons that leads to second-order equations of motion both for the scalar and the metric. 

\begin{figure}
\includegraphics[width=0.9\textwidth, natwidth=610,natheight=642]{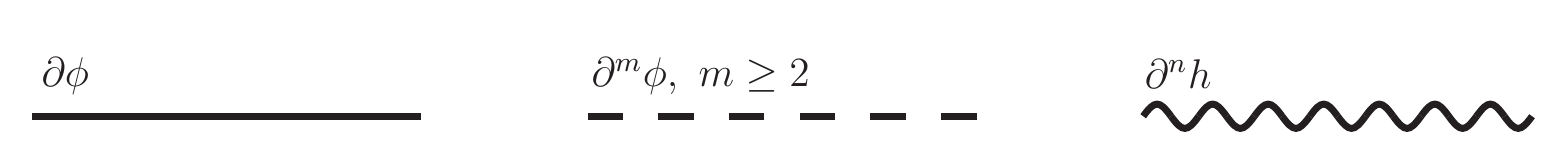}\centering
\caption{For every given vertex, a solid straight line corresponds to a possible external leg resulting in a scalar with at most one derivative acting on it. A dashed line corresponds to an external scalar with at least two derivatives, while a wiggly line denotes an external graviton.}
\label{fig1}
\end{figure}

\begin{figure}
\includegraphics[width=0.7\textwidth, natwidth=610,natheight=642]{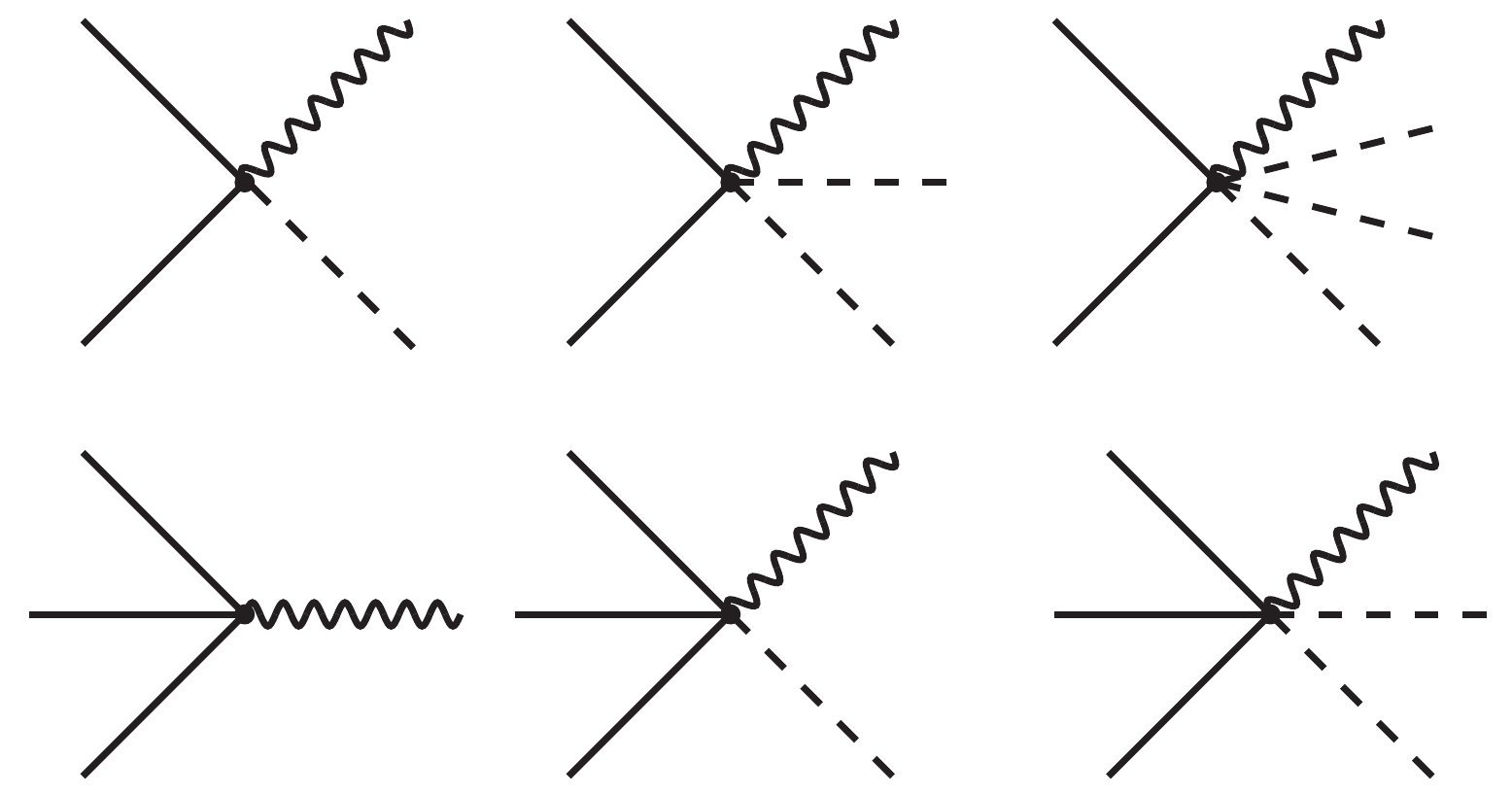}\centering
\caption{Single-graviton vertices, which can contribute two or three external scalars with one derivative ($(\p\phi)^2$ or $(\p\phi)^3$) to a 1PI vertex. The vertices from the second line exactly cancel for the unique (non-minimal) curved-space extension of the galileon that retains second-order equations of motion.}
\label{fig2}
\end{figure}

Let us now prove all of the above-described statements, starting with a few definitions. Consider an arbitrary vertex of the form $(\p\phi)^{k}(\p^m\phi)^n\p^l h^p$, with $m\geq 2$. In terms of Feynman diagrams, we will adopt the convention according to which a solid straight line corresponds to an external leg coming out of this vertex, which results in a scalar with one derivative acting on it. A dashed line corresponds to an external scalar with at least two derivatives, while a wiggly line denotes an exernal graviton, see Fig.~\ref{fig1}. For example, a $k=2,~n=1,~p=1$ vertex would correspond to the first diagram on the upper line of Fig.~\ref{fig2}. In certain cases, the number of solid lines can be less than the number of factors of $\p\phi$ in a Lagrangian interaction term. For example, the special structure of the pure galileon interactions makes them equivalent to vertices with only dashed lines, even though on average these terms contain less than two derivatives per field. This is because a galileon vertex can only lead to external states with at least two spacetime derivatives -- the fact that lies at the heart of the non-renormalization theorem associated with these theories \cite{Luty:2003vm}. We now wish to show that there exist more redundant vertices of this kind in the suitable non-minimal extension of the theory to curved space.

Let us have a look at all possible vertices with a single graviton line in the minimally coupled galileon theory, obtained from \eqref{gallag} by replacing all derivatives with covariant ones ($\p_\mu\to\nabla_\mu$). A straightforward inspection yields that there are six of such vertices, shown in Fig.~\ref{fig2}. Vertices with three solid lines can in principle arise from picking up a factor of $\p h$ from covariant derivatives\footnote{Moving the derivative on $h$ to the rest of the fields in the vertex generically does not reduce the number of solid lines for the quartic and quintic galileons, since this extra derivative can act on the factor $\p^2\phi$, representing the dashed line.}, i.e.~$\nabla^2\phi \sim \p^2\phi+\p h\p\phi$, where we denote by $\nabla^2$  any product of two covariant derivatives applied on $\phi$. 
This means that, in the minimally coupled galileon theory, the smallest scale by which the operators of the form \eqref{oneder} with $n=3 a, ~k=2 a$ are suppressed, is\footnote{This can be seen by inserting enough number of vertices with one graviton and three solid lines into a generic 1PI loop diagram.} $\sim (\mpl\Lambda_3^{5})^{1/6}$. The latter suppression scale is still too small to be consistent with the generic definition of WBG invariance, which we will introduce and discuss extensively in what follows.

We will now show that it is possible to modify the theory by non-minimal couplings to gravity that result in the elimination of vertices with three solid lines, leaving just a factor of  $(\p\phi)^2$ per graviton line. Insertion of these vertices into a generic loop diagram can only lead to symmetry-breaking operators of the form \eqref{oneder} suppressed by at least the scale $\Lambda_2 = (\mpl\Lambda_3^3)^{1/4}$, parametrically higher than $ (\mpl\Lambda_3^{5})^{1/6}$. Therefore, there is a well-defined sense in which the resulting theory is `more invariant' under the internal galileon transformations than a generic curved-space extension of the galileon. This defines what we mean by `theories with WBG invariance' throughout the present work and we will sharpen this definition in the next section, where non-trivial vacua of the resultant theories are considered.

For the cubic covariant galileon, it is easy to show that the first vertex from the lower line of Fig.~\ref{fig2} is absent: the only way to generate it is by picking up a metric perturbation from the covariant laplacian. This yields a term of the form $(\p\phi)^3\p h$. Putting the derivative from $h$ on the rest of the fields via partial integration however makes it evident that the corresponding vertex can only have two solid lines, but not three.  The case of quartic and quintic galileons is more non-trivial, but straightforward; we show in Appendix \ref{appA} that vertices with three solid lines and one graviton, as well as five solid lines and two gravitons can be removed by suitably adding non-minimal couplings to gravity.
The resultant theory is the covariant galileon of Ref.~\cite{Deffayet:2009wt} -- characterized, as a bonus, by second-order equations of motion both for the scalar and the metric. 

In summary, the `most symmetric' generalization of the galileon coupled to gravity consists of the following operators 
\begin{align}
\label{g1}
\frac{1}{\Lambda_3^3}\mathcal{L}_3 &\to \sqrt{-g} 
~\frac{1}{\Lambda_3^3}
~\mathcal{L}_3^{\text{min}}, \\
\label{g2}
\frac{1}{\Lambda_3^6} \mathcal{L}_4 &\to  \sqrt{-g} 
~ \frac{1}{\Lambda_3^6}
~\bigg [(\p\phi)^4~ R-4 ~\mathcal{L}_4^{\text{min}}\bigg ]~, \\
\label{g3}
\frac{1}{\Lambda_3^9} \mathcal{L}_5 &\to \sqrt{-g}
~\frac{1}{\Lambda_3^9}
~\bigg [(\p\phi)^4 ~G^{\mu\nu}\nabla_\mu\nabla_\nu\phi+\frac{2}{3} ~\mathcal{L}_5^{\text{min}}\bigg ]~,
\end{align}
where $\mathcal{L}_I^{\text{min}}$ denote the galileons \eqref{gal1}--\eqref{gal3} minimally coupled to gravity.
The structure of the full effective theory is such that every pair of external scalars with no more than one derivative on each unavoidably comes with a suppression of at least one power of the Planck scale. With this in mind, quantum-mechanically generated operators of the form $(\p\phi)^{2n}$ can be estimated simply on dimensional grounds to be at most of the following magnitude
\beq
\label{qcorr}
\frac{(\p\phi)^{2n}}{\mpl^n\Lambda_3^{3n-4}}~.
\eeq
The scales $(\mpl^n\Lambda_3^{3n-4})^{1/(4n-4)}$ suppressing such operators approach $\Lambda_2$ only for asymptotically large $n$, otherwise being parametrically larger. 

In the following sections, we will argue that the statement of approximate galileon invariance can be viewed as a statement about \textit{non-trivial} classical vacua, generically present in the shift-invariant theories at hand. If these vacua are to be insensitive to UV physics, the symmetry-breaking operators in the effective theory can be \textit{at most} of order \eqref{qcorr} in magnitude. In such cases we will say that the galileon invariance is broken only weakly by couplings to gravity.

\section{Theories with WBG invariance}
\label{secthree}
Perhaps, the most important lesson that one can draw from the discussion of  the previous section is that the covariant galileon is in fact \textit{not} the most general theory enjoying the above-described quantum properties. Consider the effective theory with the leading operators given by the following subclass of the Horndeski terms 
\begin{align}
\label{hor1}
\mathcal{L}^{\rm WBG}_2&=\Lambda_2^4 ~G_2(X) ~,\\
\mathcal{L}^{\rm WBG}_3&=\frac{\Lambda_2^4}{\Lambda_3^3} ~G_3(X)[\Phi] ~,\\
\label{planck}
\mathcal{L}^{\rm WBG}_4&=\frac{\Lambda_2^8}{\Lambda_3^6}  ~ G_{4}(X) R+2 \frac{\Lambda_2^4}{\Lambda_3^6}  ~G_{4X}(X)\( [\Phi]^2-[\Phi^2] \)~,\\
\label{hor4}
\mathcal{L}^{\rm WBG}_5&=\frac{\Lambda_2^8}{\Lambda_3^9} ~G_{5}(X)G_{\mu\nu}\Phi^{\mu\nu}-\frac{\Lambda_2^4}{3 \Lambda_3^9} ~ G_{5X}(X)\([\Phi]^3-3[\Phi][\Phi^2]+2[\Phi^3] \) ~,
\end{align}
where we now extend the notation employed in Eqs.~\eqref{gal1}--\eqref{gal3} by replacing the partial derivatives with covariant ones, i.e.~$[\Phi]\equiv g^{\mu \nu} \nabla_\mu \nabla_\nu\phi,~[\Phi^2]\equiv \nabla^\mu\nabla_\nu\phi\nabla^\nu\nabla_\mu\phi$, etc. 
Moreover, $G_I$ are arbitrary dimensionless functions of the dimensionless variable 
\be
X\equiv - \frac{1}{\Lambda^4_2} g^{\mu\nu}\p_\mu\phi\p_\nu\phi ~,  \label{X_def}
\ee
and the subscript `$X$' means differentiation with respect to this variable\footnote{The particular form of the interactions \eqref{hor1}--\eqref{hor4} (the relative coefficients between two operators in $\mathcal{L}^{\rm WBG}_{4}$ or  $\mathcal{L}^{\rm WBG}_{5}$, for example) may appear tuned, and this is sometimes presented as an unfortunate feature in the literature. We stress that this `tuning', motivated by restoring unitarity in the theory, is in fact natural and stable under quantum corrections. This is very similar to why we work with a gauge-invariant kinetic term $- \text{Tr}~ F^{\mu \nu} F_{\mu \nu}$ in theories with massive spin-1 particles, despite of there being no gauge invariance in the presence of the mass term.}. 
We will then assume that the functions $G_I$ are defined through the Taylor expansion
\beq
\label{gfunct}
G_I = c^{(0)}_{I}-c^{(1)}_{I} X+c^{(2)}_{I} X^2+\dots=c^{(0)}_{I}+c^{(1)}_{I} \frac{(\p\phi)^2}{\Lambda_2^4}+c^{(2)}_{I} \frac{(\p\phi)^4}{\Lambda_2^8}+\dots~,
\eeq
where $c^{(n)}_{I}$ are dimensionless, order-one coefficients. 
Note that by setting all the coefficients $c^{(n)}_{I}$ to zero except  $c^{(1)}_2 $, $c^{(1)}_{3}$, $c^{(2)}_{4}$ and $c^{(2)}_{5}$ we recover the covariant galileon of Eqs.~\eqref{g1}--\eqref{g3}. 

As in the example of the introductory section, the theory at hand is characterized by \textit{two} scales\footnote{The Planck scale arises here from ${\cal L}^{\rm WBG}_4$ in Eq.~\eqref{planck} as $\mpl^2=2 c^{(0)}_{4}\Lambda_2^8/\Lambda_3^6$.}: $\Lambda_2$ and $\Lambda_3\ll \Lambda_2$. From the EFT standpoint that we are adopting here, all Wilson coefficients $g^{(n)}_{I}$ are measured in the units of the cutoff $\Lambda_3$. The theory \eqref{hor1}--\eqref{hor4} is the one for which a well-defined class of (canonically normalized) operators -- those that respect the galileon symmetry $\phi\to\phi+b_\mu x^\mu$ -- have order-one coefficients, 
\be
g^{(n)}_{I} \sim {\cal O}(1)~, 
\ee 
while all the others are parametrically suppressed with respect to the coefficients $c^{(n)}_{I}$ by positive powers of the ratio $\Lambda_3/\mpl$. As a consequence, one can define a scaling limit
\beq
\label{declim}
\mpl\to \infty,\qquad \Lambda_2\to \infty, \qquad  \Lambda_3=\text{finite}~,
\eeq
in which the system recovers the exact internal galileon symmetry. 

A comment about the terms $c^{(1)}_{4}$ and $c^{(1)}_{5}$ is in order here\footnote{The operator $c^{(0)}_{2}$ is just the cosmological constant, which for simplicity we tune to zero, $c^{(0)}_{2}=0$. Moreover, the operators $c^{(1)}_{2}$ and $c^{(0)}_{4}$ set the normalization of the scalar and graviton kinetic terms respectively, while $c^{(0)}_{3}$ and $c^{(0)}_{5}$ give inconsequential total derivatives and can be disregarded.}. 
These terms are absent in the covariant galileon, which starts with $G_{4}, G_{5} \sim  (\p\phi)^4$. Thus, one can wonder whether they modify our conclusions regarding quantum corrections. By expanding ${\cal L}_4$ and ${\cal L}_5$ at linear order in metric perturbations, it is straightforward to verify that $c^{(1)}_{4}$ and $c^{(1)}_{5}$ do lead to new operators at the scale $\Lambda_3$. However, these operators are  just a scalar-tensor generalization of the galileon familiar from the decoupling limit of dRGT massive gravity \cite{deRham:2010ik},
\be
\begin{split}
\label{massgravityterms}
&\mathcal{L}^{\rm WBG}_4\sim h_{\mu\nu}\varepsilon_{\mu\alpha\rho\lambda}\varepsilon_{\nu\beta\sigma\lambda}
\p_\alpha\p_\beta\phi\p_\rho\p_\sigma\phi+\dots~, \\ &\mathcal{L}^{\rm WBG}_5\sim h_{\mu\nu}\varepsilon_{\mu\alpha\rho\lambda}\varepsilon_{\nu\beta\sigma\delta}
\p_\alpha\p_\beta\phi\p_\rho\p_\sigma\phi\p_\lambda\p_\delta\phi+\dots \;.
\end{split}
\ee
The latter interactions obey the same non-renormalization theorem as the galileon \cite{deRham:2012ew}, forbidding asymptotic states with less than two derivatives acting on them. In particular, this means that the counting of $\mpl^{-1}$ factors in loop diagrams of the previous section goes through unaltered: each factor of $(\p\phi)^2$ arising from a generic $1$PI vertex still comes with a suppression of at least a factor of $\mpl^{-1}$, and our analysis of quantum corrections performed for the covariant galileon also applies for the generalized theory \eqref{hor1}--\eqref{hor4}.\footnote{That the terms \eqref{massgravityterms} can be rewritten as a certain limit of  non-minimally coupled scalar-tensor theory has been noticed in \cite{Chkareuli:2011te} and the cosmology of that theory has been studied in \cite{deRham:2011by}.} Therefore, for simplicity we will disregard these terms and assume that $G_{4}$ and $G_5$ start at least quadratic in $X$. Indeed, our discussion of the previous chapter guarantees that, once tuned to zero, $G_{4X}$ and $G_{5X}$ do not receive any appreciable quantum corrections as a result of the approximate galileon symmetry.

Notice that the theories that we propose do \textit{not} reduce to the galileon for general functions $G_I$, even if one switches off gravity. Rather, they describe a certain generalization thereof, which still has no more than second-order equations of motion. The symmetry \eqref{g_inv}  is thus broken even when one sets $h_{\mu \nu}=0$ (note that this is different from taking the decoupling limit \eqref{declim}). Nevertheless, the corresponding breaking of invariance under \eqref{g_inv} is in a well-defined sense `not stronger' than the one already present once gravity is turned on. In particular, it follows from the very construction of the operators in \eqref{g1}--\eqref{g3} that each pair of factors of $\p\phi$ still comes with a suppression of at least a factor of $\mpl$ in the full effective theory. Thus, 
the Lagrangians \eqref{hor1}--\eqref{hor4}  define the most general scalar-tensor theory consistent with our definition of WBG invariance. 

The peculiar quantum properties characteristic of the theories at hand put loop corrections under complete control for a broad class of physical backgrounds. This can be established by noting that loop-generated operators always have at least one extra factor of $\mpl^{-1}$, compared to their `tree-level' counterparts. Indeed, working in the units in which $\Lambda_3=1$, each factor of $\Lambda_2^4$ becomes equivalent to $\mpl$, and the non-minimal terms in \eqref{hor1}--\eqref{hor4} can be schematically written as 
\beq
\mpl  R (\p\phi)^2 \(1+\frac{(\p\phi)^2}{\mpl}+\dots\) (\nabla^2\phi)^n~,
\eeq
while the similar loop-generated terms can never have the $\mpl$-enhancement. Likewise, the symmetry-breaking scalar operators generated quantum-mechanically are bound to have the following schematic form 
\beq
\label{loops}
\frac{\(\p\phi\)^{2n}}{\mpl^n}\(\nabla^2\phi\)^m~,
\eeq
and the analogous terms in the original action are all enhanced by at least a factor of $\mpl$ compared to this expression. 
Restoring the factors of $\Lambda_3$, we conclude that if the field expectation values are such that
\beq
\label{stability}
|X|\lsim 1, \qquad |Y|\lsim 1 ~, 
\eeq
where $X$ is defined in Eq.~\eqref{X_def} and 
\be
Y\equiv \frac{\nabla^2 \phi}{ \Lambda_3^3} \;, \label{Z_def}  
\ee
the loop corrections are negligible: they can never compete with the operators from the classical action \eqref{hor1}--\eqref{hor4} and all predictions based on the latter can be trusted in the full quantum theory. 
Moreover, not only are the magnitudes of the various operators technically natural, but any possible tuning of these can only be spoiled by corrections of higher order in $\Lambda_3/\mpl$. This generalizes the galileon non-renormalization theorem in the presence of gravity. 

On the other hand, if the conditions \eqref{stability} are not satisfied, the background becomes too strong to be trusted: the magnitude of loop corrections becomes pumped up beyond that of the leading operators and galileon invariance alone is useless in organizing the low-energy expansion. One could only trust the classical theory beyond that point if there is an extra structure, leading to a resummation of the series \eqref{gfunct}. An example would be, e.g., the DBI theories \cite{Silverstein:2003hf,Alishahiha:2004eh} (as well as generalizations thereof \cite{Hinterbichler:2012fr,Hinterbichler:2012yn}), where this happens due to non-linearly realized higher-dimensional spacetime symmetries. However,  here we want to remain as general as possible  and do not assume any extra symmetry beyond the (weakly broken) galileon one. 
In the latter case, the requirement of quantum stability expressed by Eq.~\eqref{stability} places a strong upper bound on how large the predictions for various physical quantities can be.  

\section{De Sitter vacua}
\label{secfour}

Perhaps, the best way to illustrate the physical consequences of the discussion of the previous section is to resort to a concrete example. To this end, we wish to consider the following theory
\beq
\label{full}
S=\int d^4x \sqrt{-g} ~\bigg[\frac{1}{2}\mpl^2 R -\frac{1}{2}(\p\phi)^2 -V(\phi)+\sum_{I=2}^5 \mathcal{L}^{\rm WBG}_I +\dots\bigg]~,
\eeq
where $\mathcal{L}^{\rm WBG}_I$ are the  operators defined in Eqs.~\eqref{hor1}--\eqref{hor4} of the previous chapter. We have extracted and canonically normalized the scalar and graviton kinetic terms, so that $G_{2}$, $G_{4}$ and $G_{5}$ are assumed to start \textit{at least quadratic in} $X$ (see the discussion below Eq.~\eqref{massgravityterms}), while $G_3$ can have a linear piece. In general, one can allow for a small potential, 
\beq
V(\phi)=-\lambda^3\phi +\frac{1}{2}m^2_\phi \phi^2+\dots~,
\eeq
where the parameters $\lambda$, $m_\phi$, etc. can be (naturally) arbitrarily small, since they are the only ones that break the scalar shift symmetry (which otherwise is exact even on curved space).
In the next section we will discuss the implications of a weakly broken galileon symmetry for inflation. We will thus be interested in background solutions given by an approximate de Sitter space with the Hubble parameter $H$. This section is devoted to a preliminary study of such solutions. 

For the theory \eqref{full}, the scalar equation of motion on a flat FRW background reads
\beq
\label{phieqcosm}
\frac{1}{a^3}\frac{d}{dt} \big[2 a^3 \dot\phi F(X,Z)\big]=-\frac{dV}{d\phi} ~,
\eeq
where the function $F$ is given by the following expression\footnote{Note that both $G_{4X}/X$ and $G_{5X}/X$ in this equation are finite even in the $X\to 0$ limit due to our assumption that $G_4$ and $G_5$ start at least quadratic in $X$, see the discussion below Eq.~\eqref{massgravityterms}.}
\beq
\label{f}
F(X,Z) = \frac{1}{2}+G_{2X}-3 Z G_{3X}+6 Z^2 \(\frac{G_{4X}}{X}+2 G_{4XX}\)+ Z^3\(3\frac{G_{5X}}{X}+2 G_{5XX}\)~,
\eeq
with the variable $Z$ defined so as to roughly coincide in the limit $\ddot\phi\ll H\dot\phi$ with the background value of $Y$ defined in Eq.~\eqref{Z_def}, 
\beq
\label{Z_def1}
 Z \equiv \frac{ H \dot\phi}{\Lambda_3^3}~. 
\eeq
Moreover, the two Friedmann equations can be written in the following way
\begin{align}
\label{fried1}
3\mpl^2H^2= &\ V +\Lambda_2^4X\bigg[\frac{1}{2}-\frac{G_2}{X}+2G_{2X}-6ZG_{3X}-6Z^2\(\frac{G_{4}}{X^2}-4\frac{G_{4X}}{X} -4G_{4XX} \)\nn 
\\&\qquad \qquad +2Z^3\(5\frac{G_{5X}}{X}+2G_{5XX}\)\bigg ]~, \\
\label{fried2}
\mpl^2\dot H = &-\frac{\Lambda_2^4 X F +\mpl\ddot \phi(X G_{3X}-4 ZG_{4X}-8ZXG_{4XX}-3 Z^2 G_{5X}-2Z^2XG_{5XX})}{1+2 G_{4}-4X G_{4X}-2ZXG_{5X}}~. 
\end{align}
By choosing appropriate combinations of the coefficients $c_I^{(n)}$ and in the absence of the potential, one can check that there exist \textit{exact} linear solutions $\phi_0\propto t$ to \eqref{phieqcosm} with $F(X_0,Z_0)=0$, sourcing an exact de Sitter space. Upon turning on a small enough potential for $\phi$, these backgrounds can slightly deviate from de Sitter, $\dot H \ll H^2$. Moreover, if both $X_0$ and $Z_0$ are of order one on these solutions, the background curvature is of the order $H^2\sim \Lambda_2^4/\mpl^2$ and all terms involving $\phi$ in \eqref{full} contribute equally to the energy density. 

More generally, once shift-symmetry is broken, the parameters $X$ and $Z$  measure the departure of quasi de Sitter solutions from slow-roll inflation. In particular, for $X\ll 1$ and $Z\ll 1$ one reproduces the standard potential-dominated slow-roll regime. In the opposite limit and for small enough $V(\phi)$, the acceleration can be supported by the kinetic part of the action. At the same time, we have argued below Eq.~\eqref{stability} that the same two parameters control the magnitude of quantum corrections. 
Indeed, in the $X \lsim 1$ and $Z\lsim 1$ regime loop corrections are fully under control, which can be seen by estimating typical loop-generated terms, e.g.~of the form
\beq
\label{estimates}
\frac{(\nabla^2 \phi)^n}{\Lambda_3^{3n-4}}\sim Z^n\Lambda_3^4~,\qquad \frac{(\p\phi)^{2n}}{\mpl^n \Lambda_3^{3n-4}} \sim X^n \Lambda_3^4 ~.
\eeq
For $X$ and $Z$ less than or of order one, these are parametrically suppressed with respect to the background energy density contributed by the classical Lagrangian, which, as follows from Eq.~\eqref{fried1}, generically satisfies $3\mpl^2 H^2 \gsim X\Lambda_2^4$. 
Saturating either one, or both of the bounds in Eq.~\eqref{stability} leads to a non-linear regime where various next-to-leading order terms in the derivative expansion become important, while quantum corrections are still under control. This case will be of our prime interest  in the rest of the paper. In particular, in the next section we will explore the properties of the corresponding backgrounds in the context of inflationary physics. 

\section{Inflation}
\label{secfive}

\subsection{The effective field theory for perturbations}

The consequences of approximate galileon symmetry for the physics of the early universe can be conveniently studied in the effective field theory of inflation (EFTI) framework \cite{Creminelli:2006xe,Cheung:2007st}. 
Operationally, the EFT of interest is conveniently constructed in the unitary gauge, i.e. by choosing the special coordinate system where the time slicing coincides with the uniform inflaton hypersurfaces.  The spacetime metric can be decomposed into the ADM variables,
\be
ds^2=-N^2 dt^2+\gamma_{ij}\(N^i dt+dx^i\)\(N^j dt+dx^j\) \;,
\ee
where $N$ is the laspe and $N^i$ the shift, while $\gamma_{ij}$ is the 3d metric.

The `breaking' of time diffs allows to include objects that are not invariant under the full 4d diffs, but preserve the reparametrizations of the 3d coordinates on equal-time hypersurfaces. Thus, the possible building blocks of the EFT of interest are homogeneous functions of time, intrinsic and extrinsic curvatures of equal-time hypersurfaces, respectively $^{(3)}\!R_{ij}$ and $K_{ij}$, and temporal components of tensors with upper indices such as $N=1/\sqrt{-g^{00}}$ \cite{Creminelli:2006xe,Cheung:2007st}. 
To construct the action up to quadratic order in metric fluctuations, we perturb these quantities around a flat FRW background, $ds^2 = - dt^2 + a^2(t) d \vec x^2$, by writing $\delta N \equiv N - 1$, $\delta K_{ij} \equiv K_{ij} - H \gamma_{ij}$ ($K \equiv K^i_{~i}$), while $^{(3)}\!R_{ij}$ is already at least linear in perturbations. 
In terms of these quantities, the  action reads \cite{Creminelli:2006xe,Cheung:2007st,Gleyzes:2013ooa,Piazza:2013coa}
\be
\label{s_pi}
\begin{split}
S =   \int & \ d^4x \sqrt{ \gamma }\, N \, \bigg[\frac{\mpl^2 }{2} f(t) \Big({}^{(3)}\!R+K^{ij}K_{ij}-K^2\Big) - 2 \dot f (t) \frac{K}{N}+\frac{c(t)}{N^2} - \Lambda(t)
 \\
& +\frac{M^4(t) }{2}  \delta N^2 -\hatM^3(t)\delta K \delta N -\frac{\bar M^2(t)}{2} \big( \delta K^2 - \delta K^{ij} \delta K_{ij} \big)+\frac{\tilde m^2 (t)}{2} \, {}^{(3)}\!R  \, \delta N  \\&-\frac{\bar M'{}^2(t)}{2} \big( \delta K^2 + \delta K^{ij} \delta K_{ij} \big) +m_1(t) ^{(3)}\!R\delta K+ \dots  \bigg]~.
\end{split}
\ee
The ellipses stand for Lagrangian terms constructed of space or time derivatives of $N$, $^{(3)}\!R_{ij}$ and $K_{ij}$. Moreover, we have omitted the operator ${}^{(3)}\!R_{ij}\delta K^{ij}$, since at the quadratic order it can be rewritten as a combination of other operators in \eqref{s_pi}, see \cite{Gleyzes:2013ooa}.

As far as inflation is concerned, one can always remove the time-dependent function $f$ by means of a conformal transformation of the metric \cite{Cheung:2007st}. Thus, in this section we will assume $f =1$, in which case the time-dependent  coefficients $c$ and $V$ in the first line -- those that lead to tadpoles -- are uniquely determined in terms of the background equations of motion,  
\be
c = -\mpl^2 \dot H \;, \qquad \Lambda = \mpl^2 (3 H^2+\dot H)  \;.
\ee
The rest of the action contains operators whose coefficients are in principle unconstrained.  
Depending on which subset of these coefficients is nonzero, the above effective theory describes the linear perturbations in different (and all possible) models of single-field inflation.

Given that a generic model of inflation can be defined via the set of EFT coefficients $M^4(t), \hatM^3(t), \bar M^2(t), \dots$, it is important to understand what are, if any, the possible hierarchies among these. Indeed, the basic properties of the inflationary background and perturbations such as stability, power spectra, non-gaussianities, etc., crucially depend on the relative magnitudes of these terms. In the case that the problem involves widely separated scales (such as, for instance, the Hubble scale and the UV cutoff of the theory), it is not \textit{a priori} clear what the quantum-mechanically stable values of these Wilson coefficients are. In particular, treating them as arbitrary parameters can easily lead to results that require tuning, once quantum corrections are taken into account. 

The easiest way to arrive at various hierarchies in quantum field theory is  using symmetries -- exact or approximate. For example, chiral symmetry, although inexact, is the central reason behind the technical naturalness of fermion masses in the standard model of particle physics. In the same spirit, one can use various symmetries in the effective theory \eqref{s_pi} to constrain the EFT coefficients in different models of inflation. One obvious possible symmetry are constant time translations, $t\to t+c$, which would constrain all coefficients (including the Hubble rate) in the EFT to be time-independent, enforcing the background solution to be an exact de Sitter space. Another symmetry one can impose are arbitrary space-independent reparametrizations of time, $t\to f(t)$; see, e.g.,~\cite{Blas:2010hb}. In the exact symmetry limit, this would require $M, \hatM$ and $\tilde m$ to vanish, leading to interesting theories in one corner of the inflationary theory space \cite{Creminelli:2012xb}. On the other hand, even if  time reparametrizations are not an exact symmetry, their \textit{weak} breaking (appropriately defined in what follows) would guarantee that any small values of these coefficients are technically natural. 

Building on the insights of the previous sections, we wish to investigate the consequences of yet another possible \textit{approximate} invariance of the action \eqref{s_pi} -- that induced by internal galileon transformations realized on the foliation scalar as in \eqref{g_inv}. If this scalar acquires a linear profile $\phi\propto t$ driving a de Sitter background (similarly to what we discussed in the previous section), the galileon symmetry \eqref{g_inv} will have the following manifestation in the unitary gauge 
\beq
\label{gi0}
t\to t+\tilde b_\mu x^\mu~,
\eeq 
where $\tilde b_\mu$ and $\tilde c$ are a constant four-vector and a scalar, respectively.  
For more complicated $\phi$-profiles, the associated unitary gauge invariance will have a more complicated form. However, in practice it always makes life easier to restore the foliation scalar in the effective action \eqref{s_pi} in order to explore its approximate symmetries. We will take this route in what follows.

\subsection{`Ordinary' effective theories}

Before turning to the case of models with galileon symmetry, it is instructive to recall what happens in more standard scalar effective theories. There, one starts with an action organized in derivative expansion containing a single scale $\Lambda$ -- the EFT cutoff,
\beq
\label{simplescalartheory}
S = S_{EH} + \int d^4 x \sqrt{-g}\bigg[  \Lambda_c^4 G_2\(X\) -V(\phi) +     \frac{(\Box\phi)^2}{\Lambda_c^2}+\dots\bigg ]~,
\eeq
where, as before, $G_2$ is a dimensionless function of the variable $X=-(\p\phi)^2/\Lambda_c^4$, parametrizing the leading derivative effects. 

As discussed in the previous section, for a flat enough potential the scalar background profile can be approximated by a linear function of time $\phi_0 =c(t)\Lambda_c^2 t$, where $c(t)$ is a slowly varying function. This is true both for slow-roll inflation, where $d \ln\dot \phi/dN\sim d\ln\dot H/dN$ is 
of the order of the slow-roll parameters, and for models such as the ghost condensate, where the field `velocity' is explicitly constant, i.e.~$c =$  const.~at the leading order \cite{ArkaniHamed:2003uz}. For definiteness, let us  concentrate on the latter class of models and assume that $c\sim 1$ solves $G_{2X}(c^2)=0$. On such solutions, the Hubble rate can be estimated as $\mpl^2 H^2\sim \Lambda_c^4$ and is completely determined by the leading $G_2(X)$ piece, higher-derivative operators being unimportant for the backgrounds at hand. For example, the quadratic operator shown in Eq.~\eqref{simplescalartheory} can be estimated as $(\Box\phi_0)^2/\Lambda_c^2\sim \Lambda_c^4 (H/\mpl)$ and thus it is negligible for $H\ll \mpl$, as required for a consistent classical description. 

Restricting to the unitary gauge, $\delta\phi(t,x)=0$, one straightforwardly finds that all EFT coefficients are determined by the cutoff of the theory
\beq
M^4\sim \Lambda_c^4 \;,\qquad \hatM^3\sim \Lambda_c^3 \; ,\qquad \bar M^2\sim\Lambda_c^2 \;, \qquad \bar M'{}^2\sim\Lambda_c^2\;, \qquad \text{etc.} 
\eeq
In this case, the dynamics of small perturbations is fully dominated by the only zero-derivative quadratic operator in the effective theory -- $\delta N^2$ -- and most of the phenomenology is thus determined by the single coefficient $M^4$. This leads to interesting characteristic features, such as the possibility of small speed of sound $c_s^2$ of scalar perturbations (if $|M^4|\gg |\mpl^2 \dot H|$) and, associated to it, large non-gaussianities $f_{\rm NL}\sim 1/c_s^2$~\cite{Chen:2006nt,Cheung:2007st}. The contribution of higher-order terms amounts to only slightly correcting the leading results. For example, one can show that the correction to the speed of sound from the operator $\bar M^2$ (which arises entirely due to mixing with gravity) is of order $\delta c_s^2 \sim \bar M^2 / \mpl^2\sim H/\mpl$ and can be safely ignored. Consequently, one can consistently consider the \text{perturbations} of the inflaton field   as weakly coupled over a sufficiently broad range of distances encompassing the Hubble scale and straightforwardly apply the derivative expansion. 
This would correspond to the case of \textit{k}-inflation \cite{ArmendarizPicon:1999rj}, ghost inflation  \cite{ArkaniHamed:2003uz} and other related models.

\subsection{Inflation with WBG symmetry}

\renewcommand{\arraystretch}{2.2}
\begin{table}[t]
\begin{adjustbox}{max width=\textwidth}
\begin{tabular}{|c||c|c|c|c|c|c|c|cl}
  \hline
$\mathcal{L}_I^{\rm WBG}$ &$\displaystyle {M^4}{\Lambda_2^{-4}} $ & $ \hatM^3 H  \Lambda_2^{-4} $ &$ \bar M ^{2} H^2  \Lambda_2^{-4}$&  $\tilde m^{2} H^2  \Lambda_2^{-4} $ \\
 \hline\hline
 $ I=2 $ & $4   X^2G_{2XX}$ &$\times$ &$\times$&  $\times$ \\
  \hline
$ I=3 $ &$ \displaystyle -6  X Z \left( G_{3X}+2 X G_{3XX}\right)$ &$\displaystyle -2 {  X Z} G_{3X}$ &  $\times$&  $\times$\\
  \hline 
$ I=4 $ &  $ \displaystyle 24  X Z^2\left(3G_{4XX} + 2X  G_{4XXX}\right)$&{$ \displaystyle8 { X Z^2}\left( \frac{G_{4X}}{X}+2G_{4XX}\right)$}  & $\displaystyle -4 { Z^2} {G_{4X}}$&  $\bar M^2 H^2  \Lambda_2^{-4} $ \\
 \hline
$ I=5 $ & {$ \displaystyle 2 X Z^3 \left(3\frac{G_{5X}}{X}+12 G_{5XX}+4X G_{5XXX}\right)$}&{$ \displaystyle 2 { XZ^3} \left(3 \frac{G_{5X}}{X}+2G_{5XX}\right)$} & $\displaystyle -2 { Z^3} {G_{5X}}$&  $\bar M^2 H^2  \Lambda_2^{-4} $ \\
  \hline
\end{tabular}\\
\end{adjustbox}
\caption{Contribution from the Lagrangian terms $\mathcal{L}^{\rm WBG}_I$ to various unitary-gauge operators, defined in Eq.~\eqref{s_pi}. The relation $\bar M^2=\tilde m^2$ holds for all of these. The subscript $X$ means differentiation with respect to $X$ and all derivatives are evaluated on the background. We have assumed that all \textit{background} quantities obey `slow-roll', i.e.~$d/dt\ll H$, so that terms involving derivatives of $X$ and $Z$ have been neglected. Moreover, $\mathcal{L}^{\rm WBG}_{4}$ generically contributes a weakly time-dependent piece to the effective Planck mass through the function $f(t)$ in Eq.~\eqref{s_pi}. According to our assumption about the form of $G_4$ discussed in Sec.~\ref{secthree}, this is at least of order $X^2$ for small $X$, so that its contribution can be neglected. An analogous piece coming from $\mathcal{L}^{\rm WBG}_{5}$ is suppressed by a further factor of the slow-roll parameter.   }
\label{tab1}
\end{table}

Going back to the theories characterized by WBG invariance \eqref{full},
one can straightforwardly find the values of the various coefficients in Eq.~\eqref{s_pi}, by means of reformulating the discussion of the previous section into the effective theory language \cite{Gleyzes:2013ooa}. The contribution from each Lagrangian $\mathcal{L}^{\rm WBG}_I$ to different EFTI operators is reported in Tab.~\ref{tab1}.

Interestingly, as a generic property of Horndeski theories,  only \textit{one combination} of the operators $\delta K^2$, $\delta K_{ij}\delta K^{ij}$ and ${^{(3)}\!R}\delta N$, given by the following expression
\be
\label{comb}
-  \delta K^2 + \delta K^{ij} \delta K_{ij} +  \, {}^{(3)}\!R  \, \delta N \;,
\ee
appears in the EFT action \cite{Gleyzes:2014rba}.
Thus, the parameters $\bar M^2$ and $\tilde m^2$ are related to each other as $\bar M^2=\tilde m^2$, while all the others are zero, $\bar M'^2 =  m_1 = \ldots =0$. Moreover, it follows from the discussion of Secs.~\ref{sectwo} and \ref{secthree} that this `tuning' is perfectly stable under loop corrections, as far as $X \lsim 1$ and $Z \lsim 1$. 
This implies, for instance, that $\bar M^2\gg \bar M'{}^2$ can hold in the full quantum theory due to the (weakly broken) galileon invariance. 
Moreover, the unique combination \eqref{comb} is in fact a redundant operator and can be removed by a redefinition of the metric \cite{Gleyzes:2014rba,Gleyzes:2015pma}\footnote{This redefinition corresponds to transforming into the frame where the tensor modes propagate at the unit speed of sound \cite{Creminelli:2014wna}.}. As a consequence, \textit{the single operator} $\delta N \delta K$ \textit{-- associated with the cubic galileon -- is responsible for all the differences with respect to the more familiar DBI/ghost inflation models, as far as stability and power spectra are concerned}. 

With this in mind, we derive the quadratic action for the comoving curvature perturbation $\zeta$ (defined in the standard way, $\gamma_{ij}=a^2 e^{2\zeta}\delta_{ij}$) for the general case of nonzero coefficients $M^4$ and $\hatM^3$ in Appendix \ref{appB}. The full result (away from any decoupling/short-distance limit) reads
\beq
S_\zeta = \int d^4 x~a^3 \bigg [ A(t) \dot \zeta^2-B(t)\frac{(\vec \nabla\zeta)^2}{a^2}     \bigg ]~,
\eeq
where
\ba
\label{A}
A &=& \frac{\mpl^2 (-4\mpl^4 \dot H+3 \hatM^6+2 \mpl^2 M^4)}{(\hatM^3-2\mpl^2 H)^2}, \\
\label{B}
B&=& \frac{\mpl^2 (-4\mpl^4 \dot H+2 \mpl^2 H\hatM^3 - \hatM^6+2 \mpl^2 \p_t \hatM^3)}{(\hatM^3-2\mpl^2 H)^2}~,
\ea
and the speed of sound for short-wavelength perturbations is simply $c_s^2=B/A$.
Depending on the precise values of the parameters $X$ and $Z$, there are two phenomenologically distinct regimes in a theory of inflation with WBG invariance, which we review in what follows.

\subsubsection{The kinetically driven phase}

\noindent The strongly coupled backgrounds -- i.e.~the ones for which all terms 
\eqref{hor1}--\eqref{hor4} are of the same order -- correspond to the function $F(X_0,Z_0)$ vanishing at the leading order, as already noted above. For $X$ and $Z$ strictly constant, one has an exact de Sitter space, which can be made quasi-de Sitter by turning on a small potential for $\phi$. One can see from Eq.~\eqref{f} that for $F$ to vanish requires that at least $Z_0$ be order-one. Although less evident, it is easy to show that also $X_0$ has to be order-one, if the potential is to provide a sub-leading contribution to the energy density. Indeed, it follows from Eq.~\eqref{fried1} that the energy density scales as $\mpl^2 H^2\sim \Lambda_2^4 X$ in this case. Together with the following equality, 
\beq
\label{zandx}
Z=\sqrt{X}~\frac{\mpl H}{\Lambda_2^2}~,
\eeq
which simply follows from the very definitions of $X$ and $Z$,  this implies that $Z\sim X$.
One thus concludes that whenever the theory approaches strong coupling on a (quasi) de Sitter background, $Z\sim1$, all functions $G_I$ should be generically resummed. We stress again that this only concerns the case when $V(\phi)$ can be neglected at the leading order -- a condition we will give up below.  

For large backgrounds that nearly saturate the bound \eqref{stability}, one consequence of the approximate symmetry under \eqref{g_inv} is the following \textit{radiatively stable} hierarchy 
\beq
\label{hierarchy 1}
\mpl^2 H^2 \sim M^4\sim  \hatM^3 H \sim ~ \bar M^2 H^2 \sim  \tilde m^2 H^2 ~\gg ~m_1 H^3,~  \bar M'{}^2 H^2, ~\dots \;,
\eeq 
as can be seen from Tab.~\ref{tab1}.
Here, ellipses denote all other coefficients of operators of various canonical dimension, properly normalized with the help of the appropriate number of powers of the Hubble scale.
\begin{comment}, i.e. $M_i^{4-d_i} H^{d_i}$\end{comment} 
The strongly-coupled case at hand corresponds to the largest possible values of the coefficients $M^4$, $\hatM^3$ and $\{\bar M^2,\tilde m^2\}$, i.e.~of order $\mpl^2 H^2$, $\mpl^2 H$ and $\mpl^2$ respectively. For the latter values of these operators, their effects become order-one important for physical quantities such as, for instance, the power spectrum or the speed of sound of the scalar perturbations -- see the exact expressions \eqref{A} and \eqref{B}. In this case, derivative interactions either dominate or contribute significantly to the background curvature, unlike potentially-driven models such as for instance slow-roll or DBI inflation.  
Remarkably enough, for a sufficiently large hierarchy between $\Lambda_2$ and $\Lambda_3$ -- or, equivalently, between $\mpl$ and $\Lambda_3$, the strong coupling of our backgrounds does not necessarily imply the breakdown of the classical description, as follows from the estimates of Eq.~\eqref{estimates}.  For example, in terms of the EFT language, the operator $(\Box\phi)^2/\Lambda_3^2$ leads to the $\delta K^2$ term of order $\delta\bar M^2\sim \Lambda_2^4/\Lambda_3^2\sim \mpl^2 (H/\mpl)^{2/3}$, which is suppressed by a tiny factor compared to the leading contribution\footnote{Note that quantum corrections generically introduce higher time derivatives (e.g.~$\delta \dot N^2$) at the same scales as the higher space derivatives. These lead to loss of unitarity associated with ghost excitations with masses above the EFT cutoff (cured by whatever completes the theory in the UV). The scales associated with the higher space- and time-derivative operators can only be disentangled when the speed of sound is small, $c_s\ll1$, in which case the latter scale becomes enhanced by powers of $c_s^{-1}$.}. 

There is another important characteristic to the `large' backgrounds $X\sim 1$, $Z\sim 1$ -- or, equivalently, to large values of the EFT coefficients given in Eq.~\eqref{hierarchy 1}. Assuming $\mpl^2 \dot H \lsim M^4 $, as required by being close to de Sitter space, there is no short-distance limit in which the metric can be decoupled from the inflaton perturbations. Therefore, \textit{mixing with gravity is order-one important at all scales}. To see this, let us restore the Goldstone boson $\pi$, nonlinearly realizing the spontaneously broken time diffs: $$g^{00} \to g^{00}=-N^{-2}=-1-2\dot\pi-\dot\pi^2+\frac{(\p\pi)^2}{a^2}, \qquad \delta K\to \delta K-\frac{\p^2\pi}{a^2}+\mathcal{O}(\pi^2) ~.$$ 
The relevant part of the $\pi$ action then reads
\beq
\label{piaction}
-\mpl^2 \dot H\frac{1}{N^2}+ \frac{1}{2} M^4 (\delta N)^2 -\hatM^3 \delta N\delta K\to \dot\pi_c^2+\frac{\hatM^3}{\mpl f^2_\pi}\delta N_c\frac{\p^2 \pi_c}{a^2} + \dots \;,
\eeq
where we have defined the `decay constant' $f^2_\pi \equiv ( M^4/2-\mpl^2\dot H )^{1/2}$ and canonically normalized the Goldstone and the lapse variable as $\pi=\pi_c/f_\pi^2$ and $\delta N=\delta N_c/\mpl$, respectively. One can see from the last term in this expression, that the mixing with gravity is indeed important for the Goldstone dynamics at all distance scales if the EFT coefficients saturate the strong coupling bound (i.e.~if $\hatM^3 \sim \mpl M^2$). This means that, in this case, one has to perform the full analysis including dynamical gravity in order to extract reliable predictions from the theory. That gravity is order-one important for such backgrounds can also be seen from the exact expressions for the kinetic coefficients of the curvature perturbation \eqref{A} and \eqref{B}. The calculation of the cubic action/non-gaussianities in this regime is much more involved due to the non-decoupling of gravitational effects and will be reported on elsewhere.

In principle, $Z$ may end up being somewhat smaller than one due to an accident/fine tuning (as discussed above, in our case any tuning of coefficients is respected by quantum corrections in the theories at hand). 
In this case, one can then define the decoupling limit governed solely by the dynamics of the Goldstone mode or, equivalently, by the comoving curvature perturbation $\zeta=-H\pi$. In this case the analysis beyond the quadratic action becomes significantly simpler. The  non-gaussianity in the decoupling-limit has been studied in the context of \textit{galileon inflation} in Ref.~\cite{Burrage:2010cu} and we will not reproduce the results here. Unlike DBI/ghost inflation, this limit contains cubic operators with four derivatives, such as for instance $\p^2\pi (\p\pi)^2$. However, all of these operators are redundant and can be reduced  on-shell to those with three derivatives  \cite{Creminelli:2010qf}, i.e.~$\dot \pi(\p\pi)^2$ and $\dot \pi^3$. This means that, at least in the decoupling limit, galileon inflation does not lead to new shapes of non-gaussianity, distinct from those that already arise in DBI models. The genuine novelty is that, unlike DBI theories, more than one operator becomes nonlinear in galileon inflation, leading to a richer parameter space. As a result, the amplitude of equilateral non-gaussianity $f^{\rm equil}_{\rm NL}$ is not necessarily linked to the speed of sound of scalar perturbations and it can be larger than the standard DBI result, $f^{\rm equil}_{\rm NL} \sim 1/c_s^2$, see \cite{Burrage:2010cu} for a more detailed discussion.

\subsubsection{Slow-roll inflation with WBG symmetry} 

\noindent A different and  particularly interesting case corresponds to a potentially driven universe with a subleading contribution to the energy density due to the derivative part of the action,
\beq
\mpl^2 H^2\sim V \gg \Lambda_2^4 X~.
\eeq
In the inflationary context this is just the slow-roll limit. The slow-roll parameter $\epsilon$ can be estimated from the Friedmann equations \eqref{fried1} and \eqref{fried2} to be of order
\beq
\label{epsilon}
\epsilon \equiv - \frac{\dot H}{H^2} \sim \frac{\Lambda _2^4 X}{\mpl^2 H^2}~,
\eeq
where we have made use of the fact that $F$ is an order-one function and have assumed that $F$ and $\phi$ vary slowly with time, $\dot F\ll H F,~\ddot\phi \ll H \dot\phi$. Using the identity \eqref{zandx}, the last expression for $\epsilon$ immediately implies that $Z$ has to be parametrically larger than $X$,
\beq
\label{z}
Z \sim \frac{X}{\sqrt{\epsilon}}~.
\eeq
Let us consider now a potentially-driven regime, characterized by $Z\sim 1$. According to Eq.~\eqref{z}, $X$ is parametrically suppressed, being of order $\sqrt{\epsilon}$, and it follows from \eqref{epsilon}, that
\beq
\big |\mpl^2 \dot H \big |=\mpl^2 H^2 \epsilon \sim \Lambda_2^4 X~.
\eeq 
The latter scale, denoted by $f_\pi$ in Eq.~\eqref{piaction}, defines the scalar two-point function in ordinary slow-roll models, setting the normalization for the Goldstone mode of broken time diffs,
\beq
\label{piaction1}
\mathcal{L} = -\mpl^2 \dot H (\p_\mu\pi)^2+\dots~,
\eeq 
where ellipses denote all sorts of additional terms, completely unimportant in slow-roll inflation, including those describing the mixing of $\pi$ with gravity. 

Let us now have a look at the values of the additional operators present in the case of interest and listed in Tab.~\ref{tab1}. A crucial observation is that, when $Z$ is of order one, the contribution to $M^4$ and $\hatM ^3$ from all $\mathcal{L}^{\rm WBG}_I$ with $I\neq 2$  scales as $M^4 \sim \Lambda_2^4 X\sim \mpl^2\dot H$ and $\hatM^3\sim \mpl^2 \dot H /H$ respectively. The $M^4$ operator that arises from $\mathcal{L}^{\rm WBG}_2$ is down by an additional factor of $X\sim \sqrt{\epsilon}$ compared to this, so that, as long as perturbations are concerned, higher-derivative terms are more important than the next-to-leading order terms in the derivative expansion of the form $(\p\phi)^{2n}$. Such a scaling of extra EFT operators is quite remarkable, since it implies that the perturbation Lagrangian \eqref{piaction1} is modified at order one compared to the usual slow-roll case, while the background evolution is still fully governed by the potential\footnote{This is analogous to what occurs in DBI inflation \cite{Silverstein:2003hf}, even though there are crucial differences. Here we have in mind a scenario closer to slow-roll inflation than DBI models, in that e.g.~the Hubble friction is important for the inflationary dynamics of $\phi$ and sustaining inflation requires a flat enough potential -- or, in other words, that the usual relation $\epsilon \sim \(\mpl V'/V\)^2$ holds. This is not necessarily true in DBI inflation.}.

This can also be seen from the exact expressions for the quadratic scalar perturbation Lagrangian. Taking into account the magnitudes of the two EFTI operators at hand, let us denote
\beq
\label{Ms}
M^4 = -\alpha \mpl^2\dot H = \alpha \mpl^2 H^2\epsilon,  \qquad \hatM^3 = \beta \mpl^2\frac{\dot H}{H} = -\beta \mpl^2 H\epsilon~,
\eeq
where $\alpha$ and $\beta$ are constants of order one. Then, from Eqs.~\eqref{A} and \eqref{B} one finds
\beq
A =\bigg(1+\frac{\alpha}{2}\bigg) \epsilon \mpl^2+\mathcal{O}(\epsilon^2) ~,\qquad B =\bigg(1-\frac{\beta}{2}\bigg)\epsilon \mpl^2+\mathcal{O}(\epsilon^2) \;, \qquad c_s^2 = \frac{2-\beta}{2+\alpha}~.
\eeq
This explicitly shows that, by suitably choosing the order-one coefficients $\alpha$ and $\beta$, which themselves are functions of the free parameters of the theory, the speed of sound of the scalar perturbations end up anywhere in the allowed region $0 \le c_s^2\leq 1$. Moreover, tuning $\beta$ to be close to $2$ yields a regime with highly subluminal scalar perturbations, potentially implying strong coupling and large `accidental' non-gaussianities \cite{Upcoming}, all in a technically natural way (recall that any tuning in the theories at hand is protected by WBG invariance).

Last but not least, we note that in the potentially driven regime, one can generically define the decoupling limit -- unlike the kinetically dominated case considered above. This can be seen by noting that the term describing the mixing of the (canonically normalized) Goldstone $\pi_c$ with gravity is parametrically suppressed on backgrounds close to de Sitter. Indeed, using the expressions for the magnitudes of EFT coefficients in Eq.~\eqref{Ms}, one can readily evaluate the coefficient of the mixing term in Eq.~\eqref{piaction} to be of order
\beq
\frac{\hatM^3}{\mpl f_\pi^2} \sim  \sqrt{\epsilon}~.
\eeq
During inflation, this is a parametrically small number meaning that the Goldstone mode can be studied independently from the gravitational degrees of freedom, at frequencies of order of the Hubble scale. The same conclusion can be made regarding all other EFT operators present in the theory.

One can also see from Table \ref{tab1} that, while generalizing galileon inflation, the most general inflationary theory defined by \eqref{full}  possesses a much broader parameter space. While three parameters in Tab.~\ref{tab1} characterize galileon inflation ($G_{3X}$, $G_{4XX}$, $G_{5XX}$), a generic theory with WBG invariance has six more parameters\footnote{Assuming $G_{4X}$ and $G_{5X}$ are zero, which we did in this paper.}  at the level of the two-point scalar function  ($G_{4X},~G_{5X},~G_{2XX},~G_{3XX},~G_{4XXX},~G_{5XXX}$). Higher-point functions will introduce even more freedom.
Moreover, arbitrary tuning of the parameters is stable under radiative corrections and taking advantage of this fact can lead to interesting theories in various corners of the EFT of inflation theory space. The complete analysis of possible physical predictions of the theory  \eqref{full} is beyond the scope of the present discussion.

\section{Late-time universe}
\label{secsix}
\subsection{Dark energy}

The theories characterized by WBG symmetry can also be applied in the context of the late-time cosmic acceleration. 
For simplicity, we will neglect the scalar potential $V(\phi)$ in this section, contrarily to the case discussed in the previous section. Moreover, we reabsorb the canonical scalar kinetic term in the definition of the function $G_2$, so that the action of interest reads as follows
\beq
\label{lt_full}
S=\int d^4x \sqrt{- g}  \left[ \frac{1}{2} \mpl^2 R +  \sum_{I=2}^5  \mathcal{L}^{\rm WBG}_I \right] ~.
\eeq
The homogeneous equations on a flat FRW background obtained by varying this action are similar to those, discussed in Sec.~\ref{secfive}. However, in this section we choose to rewrite them in a slightly different form and notation.

In terms of the two variables $X$ and $Z$ defined in Eqs.~\eqref{X_def} and \eqref{Z_def1}, the homogeneous scalar evolution equation reads \cite{Kobayashi:2010cm}
\beq
\frac{d}{dt} \left[ a^3\dot\phi F_{\rm DE}(X,Z) \right] =0 \;, \label{EOM_bg}
\eeq
where $ F_{\rm DE}(X,Y)$ 
is defined as
\be
F_{\rm DE}(X,Z) \equiv G_{2X} -3   Z G_{3X} + 6 Z^2 \left( \frac{G_{4X}}{X} + 2 G_{4XX} \right) + Z^3 \left(3 \frac{G_{5X}}{X} + 2  G_{5XX} \right)  \;.
\ee
We assume that all matter couples minimally to the metric $g_{\mu \nu}$, so that the Friedmann equations can be written as
\begin{align}
3 M_*^2 H^2 &=  \rho_{\rm m} + \rho_{\rm DE} \;, \\
- 2 M_*^2 \dot H & =  \rho_{\rm m} + p_{\rm m} + \rho_{\rm DE} + p_{\rm DE} \;,
\end{align}
where $\rho_{\rm m}$ and $p_{\rm m}$ respectively denote the energy density and pressure of matter. The effective Planck mass squared, $M_*^2$, is generically time-dependent due to the non-minimal couplings of the scalar in $\mathcal{L}^{\rm WBG}_4$ and $\mathcal{L}^{\rm WBG}_5$, and is explicitly given by the following expression
\beq
M_*^2 \equiv \mpl^2 \left(1 + 2 G_4 - 4 X G_{4X} - 2  ZX G_{5X}  \right)\;.
\eeq
In terms of the EFT parameters of Eq.~\eqref{s_pi}, $M^2_* = \mpl^2 f + \bar M^2$.
Finally, the unperturbed  effective energy density and pressure read
\begin{align}
\rho_{\rm DE} = & \ \Lambda_2^4 X\left[ - \frac{G_2}{X} + 2  G_{2X} - 6 Z  G_{3X} \right. \nonumber \\
& \left. + 12 Z^2 \(\frac{G_{4X}}{X} +  2  G_{4XX} \) + 4 Z^3 \(\frac{G_{5X}}{X} +   G_{5XX} \) \right] \;, \label{rhoDE} \\
p_{\rm DE} +\rho_{\rm DE}= &\ 2 \Lambda_2^4 X F_{\rm DE} + 2 \mpl  \ddot \phi X 
\nonumber \\
&  \times \left[  G_{3X}  - 4 Z \left( \frac{G_{4X}}{X} + 2  G_{4XX} \right)  - Z^2 \left(3 \frac{G_{5X}}{X} + 2  G_{5XX} \right)  \right] \;. \label{pDE}
\end{align}

For Horndeski theories, the phenomenological deviations from the $\Lambda$CDM model at the level of linear perturbations can be conveniently parametrized in terms of four time-dependent parameters \cite{Gleyzes:2013ooa}. It is convenient to use the  dimensionless functions introduced in \cite{Bellini:2014fua}, which
can be expressed in terms of the EFT coefficients of Eq.~\eqref{s_pi} as \cite{Gleyzes:2014rba}
\begin{equation}
\alpha_K =  \frac{2 c +  M^4}{M_*^2 H^2} \; ,   \qquad \alpha_B  = \frac{\mpl^2 \dot f - \hatM^3}{ 2 M_*^2 H}   \;,   \qquad
\alpha_M  =  \frac{\mpl^2 \dot f +  \dot{ \bar M}^2}{M_*^2H }  \;,  \qquad
\alpha_T  = - \frac{  \bar M^2}{M_*^2 }  \;, \label{alphaEFT}
\end{equation}
where the function $c$ is fixed by the background Friedmann equations derived from action \eqref{s_pi},
\be
2 c = p_{\rm DE} +\rho_{\rm DE} +  \mpl^2 (H \dot f - \ddot f) \;.
\ee
For the theory of interest \eqref{lt_full}, their explicit expressions  are reported in App.~\ref{appC}. The contributions from the Lagrangians ${\cal L}_I^{\rm WBG}$ to each of these parameters is given in Tab.~\ref{tab2}.
\renewcommand{\arraystretch}{2.2}
\begin{table}[t]
\begin{adjustbox}{max width=\textwidth}
\begin{tabular}{|c||c|c|c|c|c|c|c|cl}
  \hline
$\mathcal{L}_I^{\rm WBG}$&${M_*^2 H^2}{\Lambda_2^{-4}} \alpha_K $ & $ M_*^2 H^2 \Lambda_2^{-4} \alpha_B $ &$ M_*^2 H^2 \Lambda_2^{-4} \alpha_M$&  $M_*^2 H^2 \Lambda_2^{-4} \alpha_T $ \\
 \hline\hline
\rule{0pt}{3ex} 
$ I=2 $ & $2X ( G_{2X} + 2 X G_{2XX})$ &$\times$ &$\times$&  $\times$ \\
  \hline
$ I=3$ &$ \displaystyle -  12 X  Z (G_{3X} +  X G_{3XX})$ &$\displaystyle X Z  G_{3X}$ &  $\times$&  $\times$\\
  \hline 
$I=4 $ &  $ \displaystyle 12 X Z^2  \left(\frac{G_{4X}}{X} +8  G_{4XX} + 4X G_{4XXX} \right)$&{$ \displaystyle -4X Z^2 \left( \frac{G_{4X}}{X}+2G_{4XX}\right)$}  & $\displaystyle - \frac{ \ddot \phi  }{\Lambda_3^3}  \alpha_B $&  $\displaystyle 4 Z^2 {G_{4X}}$ \\
 \hline
$ I=5 $ & {$ \displaystyle 4 X Z^3  \left(3 \frac{G_{5X}}{X} + 7 G_{5XX} + 2 X G_{5XXX} \right)$}&{$ \displaystyle - X Z^3  \left(3 \frac{G_{5X}}{X}+2G_{5XX}\right)$} & $\displaystyle  2\frac{\dot H Z^3}{H^2 }  {G_{5X}} - 2\frac{ \ddot \phi  }{\Lambda_3^3}  \alpha_B $&  $\displaystyle 2 Z^2 \bigg( Z  - \frac{  \ddot \phi }{\Lambda_3^3 }  \bigg) {G_{5X}}$ \\
  \hline
\end{tabular}\\
\end{adjustbox}
\caption{Contribution from the Lagrangian terms $\mathcal{L}^{\rm WBG}_I$ to the various effective  $\alpha$-parameters defined in Eq.~\eqref{alphaEFT},  denoting the deviations from the $\Lambda$CDM model in linear perturbation theory. Their complete expressions are reported in App.~\ref{appC}, see Eqs.~\eqref{alphaK}--\eqref{alphaT}.    }
\label{tab2}
\end{table}

Let us first briefly comment on the physical meaning of the $\alpha$-parameters and their observational consequences (see, e.g.~\cite{Bellini:2014fua,Gleyzes:2014rba} for a more detailed discussion).
\begin{itemize}
\item The first of these functions, $\alpha_K$, parametrizes the kinetic energy of  scalar perturbations, induced by the four Lagrangian terms $\mathcal{L}^{\rm WBG}_I$. In terms of the  EFT for perturbations \eqref{s_pi}, the contributions to $\alpha_K$ arise from the operators $\propto \delta N^2$, described by the coefficients $c$ and  $M^4$. 
This parameter is enough to describe linear perturbations in the minimally coupled quintessence and $k$-essence models. In the minimal case, dark energy fluctuations behave as those of a perfect fluid with the speed of propagation $ c_s^2= 3 (1+w_{\rm DE} ) \Omega_{\rm DE} /\alpha_K$, where $\Omega_{\rm DE}$ is the energy density of $\phi$ in the units of the critical one.   
A well-studied example corresponds to the limit $\alpha_K \gg 1$, which leads to  zero sound speed  \cite{Creminelli:2008wc,Creminelli:2009mu}. 

\item The second function, $\alpha_B$, parametrizes the mixing between metric and scalar field fluctuations \cite{Deffayet:2010qz}. It can be induced by the Lagrangians $\mathcal{L}^{\rm WBG}_I$ with $ 3\le I \le 5$, or by  the operator containing $\delta N \delta K$ in the EFT for perturbations, which describes kinetic mixing  with gravity. This operator induces a typical scale (approximately given by $k_B/a \simeq 3  H \sqrt{\Omega_{\rm m}  / 2} $ for $\alpha_M  = \alpha_T=0$) \cite{Bellini:2014fua},  below which dark energy can cluster with energy density fluctuations $\delta \rho_{\rm DE} \simeq 2 \delta \rho_{\rm m} \alpha_B^2/  \( (\alpha_K^2 + 6 \alpha_B^2) c_s^2 \)$. Above this scale, dark energy fluctuations simply behave as  those of a perfect fluid.

\item The function $\alpha_M$ is defined as  $\alpha_M \equiv d \ln M_*^2 /\ln a$, and is thus related to the time variation of the effective Planck mass $M^2_*$ induced by the Lagrangians $\mathcal{L}^{\rm WBG}_4$ and $\mathcal{L}^{\rm WBG}_5$. It parametrizes the non-minimal coupling in scalar-tensor theories such as Brans-Dicke \cite{Brans:1961sx} and, as such, induces a slip between the gravitational potentials.
In terms of the EFT operators, $\alpha_M$ can be generated by time variation of the parameters $f$ and $\bar M ^2$. In particular, in $f(R)$ theories, $\alpha_M = - \alpha_B$. 

\item Finally, the last function, $\alpha_T$, parametrizes the deviation of the speed of sound of tensor perturbations from unity in the presence of the Lagrangian $\mathcal{L}^{\rm WBG}_4$ and $\mathcal{L}^{\rm WBG}_5$ \cite{Kobayashi:2010cm}, and also induces a slip between the gravitational potentials. It is generated by the EFT operator, multiplied by the coefficient $\bar M^2$.  
\end{itemize}

A particularly interesting case corresponds to a constant $X$, where the scalar field grows linearly with time, $\phi_0 \propto t$. In this case, the functions $G_I$, and all their derivatives are time-independent when evaluated on the background, while $Z$ changes proportionally to the Hubble rate. One can see from the form of the scalar equation, Eq.~\eqref{EOM_bg}, that the $\phi_0 \propto t$ profile can only be a solution if the following relations are satisfied
\beq
G_{2X} = 0 \;, \qquad G_{3X} = 0 \;, \qquad G_{4X} + 2 X G_{4XX} = 0 \;, \qquad 3 G_{5X} + 2 X G_{5XX} = 0\;. \label{cond1}
\eeq
It then follows from Eqs.~\eqref{rhoDE} and \eqref{pDE} that the effective energy density on such backgrounds is given by $\rho_{\rm DE} = - \Lambda_2^4 (G_2 + 2 Z^3 G_{5X})$ and the equation of state is that of the cosmological constant, with $ w_{\rm DE} \equiv p_{\rm DE} / \rho_{\rm DE} =-1$. 
Moreover, while the dimensionless parameters $\alpha_K$, $\alpha_M$ and $\alpha_T$ in Tab.~\ref{tab2} are generally nonzero and time dependent, Eq.~\eqref{cond1} implies that $\alpha_B =0 $. In this case, in order to avoid ghost instabilities one must require that $\alpha_K \ge 0$ \cite{Gleyzes:2013ooa,Bellini:2014fua}.
Note that since $Z \propto H$, $\rho_{\rm DE}$ is time dependent, even though $w_{\rm DE} =-1$. This is because the effective Planck mass varies with time and the dark energy does not follow the usual conservation equation. Demanding that the energy density remains finite at early times requires it to be constant, i.e.~$G_{5X}=0$ and thus that also $\alpha_M=0$, see Tab.~\ref{tab2}. In this case, at the background level this model exaclty behaves as a cosmological constant. At the level of linear perturbations, the dimensionless parameters $\alpha_K $ and  $\alpha_T$ do not vanish and contribute to the sound speed of scalar and tensor fluctuations, respectively as $c_s^2 = - 2 \alpha_T/\alpha_K$ and $c_T^2 = 1 + \alpha_T$ (see \cite{Bellini:2014fua,Gleyzes:2014rba}). One can avoid gradient instabilities by requiring the positivity of both these speed of fluctuations squared, which implies $-1 \le \alpha_T \le 0$.

This is the simplest application of WBG symmetry to the late time-acceleration. The assumption that  the background profile of the field is linear with time, $\phi_0 \propto t$, and that the background expansion history is the same as in $\Lambda$CDM considerably restricts the values that the parameters $\alpha_a$ can consistently  take. 
These solutions are not possible for the covariant galileon \cite{Deffayet:2009wt}. In this case one usually assumes a tracker  solution with $Z =$ const.~\cite{DeFelice:2010pv}; this imposes a particular expansion history, which has been shown to lead to observations that are disfavoured with respect to $\Lambda$CDM (see \cite{Barreira:2014jha}  for a recent analysis and e.g.~\cite{Appleby:2011aa,Neveu:2013mfa} for analysis that do not assume the tracker solution). More sophisticated examples can be constructed using the Lagrangians \eqref{hor1}--\eqref{hor4}, by allowing the background solution for $\phi$ to be different from a profile linear in time. In this case, from Tab.~\ref{tab2} one generically expects that the parameters $\alpha_a$ may assume any value smaller than unity, $\alpha_a \sim X \lsim 1$.

\subsection{Static sources and the Vainshtein mechanism}

Any theory with gravitationally coupled scalars should be able to pass the solar system constraints in order to be able to qualify as an acceptable alternative to general relativity. In the context of theories with WBG invariance, one can imagine coupling  $\phi$ to the trace of the matter stress tensor in the following way
\beq
\mathcal{L}_{mat}\sim \frac{1}{\mpl }\phi T_{\rm m}~.
\eeq
It is then necessary to make sure that the exchange of $\phi$ by realistic sources does not modify the Newtonian potential by any detectable amount. 
There are a number of ways to suppress the contribution from $\phi$ to the gravitational potential of astrophysical objects (stars, planets, etc.) in modified gravity. In the case of the theories with approximate galileon invariance we are interested in, what guarantees that $\phi$ is screened beyond the observable values is the Vainshtein mechanism \cite{Vainshtein:1972aa,Deffayet:2001uk}. Indeed, in the decoupling limit \eqref{declim}, the theory reduces to the Galileon, which is known to naturally incorporate the screening below the Vainshtein redius \cite{Nicolis:2008in}. For a souce of mass $M_{\rm source}$, this radius reads
\beq
r_V=\(\frac{M_{\rm source}}{\mpl \Lambda^3_3}\)^{1/3}~.
\eeq
The decoupling limit, on the other hand, does capture all relevant astrophysical scales and all results obtained in that limit can be fully trusted. To see this directly, one can evaluate the quantity $X$ on a pure Galileon solution of Ref. \cite{Nicolis:2008in} and check that it is extremely small everywhere in space\footnote{In the case of the cubic Galileon, $X$ becomes of order one only at the Schwarzschild radius of the source $r_S$, while, in the most general case, it is a constant within the Vainshtein radius, of order of the tiny ratio $(r_S H_0)^{2/3}$, where $H_0$ is the current Hubble rate.}, so that all the results of Ref. \cite{Nicolis:2008in} apply to our case without any modification.

\section{Conclusions and outlook}
\label{secseven}
Galileon-symmetric scalar theories have played an important role in cosmological model-building in the context of both the early and the late-time universe. The exact symmetry is only possible in the absence of gravity and it leads to remarkable properties -- both at  the classical and quantum levels. However, upon inclusion of gravity galileon invariance necessarily has to be broken, raising the question regarding the fate of its attractive flat-space properties. 

In this paper, we have introduced the notion of \textit{weakly broken galileon invariance}, which allows to naturally generalize the galileon to curved space (see Sec.~\ref{sectwo}). Requiring that the quantum non-renormalization properties of the flat-space theory is preserved to the maximal possible extent also in the presence of gravity has led us, quite uniquely, to a particular sub-class of the ghost-free Horndeski theories, presented in Sec.~\ref{secthree}. The resulting effective theory is characterized by two scales -- or, more appropriately, by parametrically small coefficients of symmetry-breaking operators in the units of the effective field theory cutoff -- allowing to retain the quantum non-renormalization properties of the galileon for a broad range of physical backgrounds. 

As we have shown, the latter properties guarantee that de Sitter solutions that these theories generically possess are insensitive to loop corrections. This gives a lot of freedom to implement these solutions for building technically natural models of both the inflationary, as well as the late-time universe. One of the purposes of this work has been to introduce these models and set the stage for their detailed phenomenological studies.

In Sec.~\ref{secfive}, we have studied two relevant regimes in the context of inflation. The first one describes genuinely strongly coupled physics, characterized by $X = \dot \phi^2 /\Lambda_2^4 \sim 1$  and $ Z = H \dot \phi/ \Lambda_3^3\sim 1$ in the notation of Sec.~\ref{secthree}. It generalizes \textit{galileon inflation} of Ref.~\cite{Burrage:2010cu} and saturates the bound on the `strength' of the backgrounds for which the classical analysis can still be trusted. The dynamics of perturbations on these backgrounds is completely dominated by higher-derivative operators. Yet, the mechanism is fully predictive, quantum loops giving subleading corrections to the classical results. In a generic theory with weakly broken galileon invariance, there are six additional parameters (two of which have been set to zero in this paper) compared to the particular case of galileon inflation at the level of the scalar two-point function. More freedom arises when higher $n$-point functions are concerned. 

The other, particularly interesting, regime corresponds to partially strongly coupled backgrounds, with $Z\sim 1$ and $X\sim \sqrt{\epsilon}$, resembling slow-roll inflation. In this scenario, inflation happens mostly due to the potential; however, weakly broken galileon invariance makes it possible for the higher-derivative operators to become relevant for the dynamics of scalar \textit{perturbations}, modifying their action at order one. In particular, this allows for a significantly subluminal speed of sound for the curvature perturbation $\zeta$ as a result of a possible parameter tuning -- which is, importantly, respected by loop corrections to the leading order in $\mpl^{-1}$. The small speed of sound generically translates into strong coupling of scalar perturbations, leading to large non-gaussianities. Such `accidentally' moderate/strong non-gaussian perturbations can never arise in ordinary slow roll models. Thus,  this represents an interesting possibility that the theories with weakly broken galileon invariance can offer. These matters will be studied in detail in an upcoming work \cite{Upcoming}.

The theories introduced in Sec.~3 can also be  used to drive the current accelerated expansion of the universe, as discussed in Sec.~\ref{secsix}. The situation is richer than the covariant galileon case studied in the literature. In particular, while one can recover a background expansion history closed to the one currently observed, at the level of linear perturbations these theories can lead to modifications of gravity of different types. Expressed in terms of the dimensionless  effective field theory parameters $\alpha_a$ defined for Horndeski theories, one  generically finds that these cannot be too large, $\alpha_a \lsim1$, leaving the possibility of modifications which are compatible with  current constraints but possibly testable by future observations. We leave a more  thorough  study of  concrete scenarios for the future.

As a final remark, we note that we have not carried out a complete analysis of quantum corrections in the theories with weakly broken galileon invariance; we have rather made the most conservative estimates of their magnitudes, so that the statements made above should remain true for the most general completion of these theories above the scale $\Lambda_3$. Whether or not our findings regarding the structure of the low-energy effective theory \eqref{full} imply anything useful about its UV completion, is still to be seen. We postpone the detailed study of this question for future work.

\subsection*{Acknowledgements}
It is a pleasure to thank Paolo Creminelli and Greg Gabadadze for valuable discussions. The work of D.P. is supported in part by MIUR-FIRB grant RBFR12H1MW and by funds provided by {\it Scuola Normale Superiore} through the program {\it "Progetti di Ricerca per Giovani Ricercatori"}. The work of E.T. is supported in part by MIUR-FIRB grant RBFR12H1MW. F.V. acknowledges kind hospitality by {\it Scuola Normale Superiore}, where part of this project was completed.
\appendix

\section{Quantum corrections}
\label{appA}

In this appendix, we extend the discussion of Sec.~\ref{sectwo} to show that the vertices  with three solid lines and one graviton, as well as five solid lines and two gravitons (see Fig.~\ref{fig2}) can be removed in a suitable curved-space extension of the galileon terms.

We have considered the case of the cubic galileon in the main text. One has to work a little more to understand the case of quartic minimally coupled galileon. The contribution to the second vertex from the lower line of Fig.~\ref{fig2} can be extracted by picking up a factor of $\p h$ from the covariant derivative acting on one of the scalars. The relevant term reads,
\beq
\label{cd}
\nabla_\mu\nabla_\nu\phi \sim -\frac{1}{2}\(\p_\mu h_{\rho\nu}+\p_\nu h_{\rho\mu}-\p_\rho h_{\mu\nu}\) \p_\rho\phi~.
\eeq
Here (and in the rest of the present section), by `$\sim$' we mean `equals up to a total derivative and up to terms with fewer factors of $\p\phi$', and we do not distinguish between upper and lower indices for simplicity (everything is contracted with the flat-space metric). Using \eqref{cd}, we find after a little bit of algebra 
\beq
\label{contr1}
-4 \mathcal{L}_4^{\text{min}}\sim - 4 (\p\phi)^2 h_{\mu\nu} \p_\rho\phi \(\p_\mu\p_\nu-\Box \eta_{\mu\nu}\)\p_\rho\phi~,
\eeq
where $\mathcal{L}_4^{\text{min}}$ denotes the Lagrangian term, obtained by minimally covariantizing \eqref{gal2}. Inserting the latter vertex into a generic $1$PI loop diagram would induce galileon symmetry-breaking operators at the scale $\mpl^{1/6} \Lambda_3^{5/6}$, parametrically exceeding $\Lambda_2$ -- as discussed above\footnote{Although parametrically higher than $\Lambda_3$, this symmetry-breaking scale is still low enough to disrupt the de Sitter spectrum \eqref{hierarchy 1}.}. Fortunately, it turns out possible to raise it by adding non-minimal couplings, capable of eliminating the vertices with three solid lines coming from the minimally coupled theory. For the quartic galileon, the right coupling is $\sqrt{-g}(\p\phi)^4R$ . Indeed, expanding the Ricci scalar to the linear order in $h$ and integrating by parts, we obtain
\beq
\label{contr2}
\sqrt{-g}(\p\phi)^4 R \sim  4 (\p\phi)^2 h_{\mu\nu} \p_\rho\phi \(\p_\mu\p_\nu-\Box \eta_{\mu\nu}\)\p_\rho\phi+\mathcal{O}(h^2)~,
\eeq
implying that the two contributions to the possible vertex with one graviton and three solid lines exactly cancel for our resulting generalized theory. 

The way we have chosen to couple the galileon to gravity has been dictated by our desire to raise the scale suppressing loop-generated symmetry-breaking operators. However, there is something more to it: the resulting non-minimal theory can be recognized as that of Horndeski type, leading to second-order equations both for the metric and the scalar. In the retrospect, this is not that surprising: the `three solid line' vertices, if present, are the only ones that would introduce higher-order equations for the metric in the theory of minimally coupled galileon, as can be easily seen from e.g. Eq.~\eqref{contr1}. Our non-minimal term exactly cancels this contribution. We thus observe something similar to what happens in massive gravity \cite{deRham:2010ik}: there, eliminating terms that lead to higher-order equations (or, equivalently, to the Boulware-Deser ghost) automatically raises the quantum cutoff of the theory; in our case, the cutoff ($\Lambda_3$) remains unaltered, but a similar procedure raises the scale of breaking for the galileon symmetry. 

The same result can be straightforwardly extended to the case of the quintic galileon. A brute-force expansion of the minimally-coupled term, using \eqref{cd}, yields
\be
\begin{split}
\label{contr3}
\frac{2}{3}\mathcal{L}_5^{\text{min}}\sim & \ (\p\phi)^2\p_\rho\phi \big[ \p_\rho h\((\Box\phi)^2-(\p_\mu\p_\nu\phi)^2\)-2\p_\rho h_{\mu\nu}(\p_\mu\p_\nu\phi \Box\phi-\p_\mu\p_\alpha\phi \p_\nu\p_\alpha\phi) \\
&- 2 \p_\mu h_{\rho\mu}\((\Box\phi)^2-(\p_\alpha\p_\beta\phi)^2\)+4 \p_\mu h_{\rho\nu}(\p_\mu\p_\nu\phi \Box\phi-\p_\mu\p_\alpha\phi \p_\nu\p_\alpha\phi)\big]~.
\end{split}
\ee
It is impossible to manipulate the last expression by partial integration or otherwise, so as to be left with less than three factors of $\p\phi$. This would lead to the last vertex with three solid lines in Fig.~\ref{fig2}, and therefore higher-order equations for the metric. One can however show that this term can be eliminated by adding a non-minimal piece with an exact coefficient corresponding to the Horndeski theory. 
To this end, we note that to the linear order in the metric perturbation, the following relation holds
\beq
\sqrt{-g}(\p\phi)^4 G_{\mu\nu}\nabla^\mu\nabla^\nu\phi = -\frac{1}{2}(\p\phi)^4\epsilon_{\mu\alpha\rho\lambda} \epsilon_{\nu\beta\sigma\lambda}\p_\alpha\p_\beta h_{\rho\sigma}\p_\mu\p_\nu\phi+\mathcal{O}(h^2)~,
\eeq
where $\epsilon$ denotes the totally antisymmetric symbol with 
$\epsilon_{1234}=1$. Expanding the antisymmetric product and integrating by parts, we obtain
\beq
\label{contr4}
(\p\phi)^4 G_{\mu\nu}\nabla^\mu\nabla^\nu\phi \sim (\p\phi)^2 \p_\rho\phi \big[\p_\rho h\((\p_\mu\p_\nu\phi)^2-(\Box\phi)^2\)+2 \p_\rho h_{\mu\nu}\(\p_\mu\p_\nu\phi \Box\phi-\p_\mu\p_\alpha\phi \p_\nu\p_\alpha\phi\)\big].\nn
\eeq
This exactly cancels the first line of the minimal term's contribution \eqref{contr3}. Partially integrating the second line of \eqref{contr3} on the other hand, gives
\beq
 -2(\p\phi)^2\p_\rho\phi \p_\nu h_{\rho\mu}\big[\p_\mu\p_\alpha\phi\p_\nu\p_\alpha\phi-\p_\mu\p_\nu\phi
\Box\phi+\p_\mu\Box\phi\p_\nu\phi-\p_\mu\p_\nu\p_\alpha\phi\p_\alpha\phi  \big]~.
\eeq
One can now check, that whatever stands to the right of $(\p\phi)^2\p_\rho\phi$ in the last expression, is a total derivative ($\p_\nu$ of a local operator). This means that one final partial integration can get rid of one solid line in the corresponding vertex, reducing the number of solid lines to two. This proves our statement regarding elimination of vertices with one graviton and more than two solid lines.

One last statement we wish to prove regards the absence of vertices with two gravitons and more than four solid lines. The only possible obstruction comes from the non-minimal coupling obtained above, $(\p\phi)^4 G_{\mu\nu}\nabla^\mu\nabla^\nu\phi$, in which we take the Einstein tensor linear in the metric perturbation, and pick up another factor of $h$ from the covariant derivative\footnote{A vertex with more than four solid lines coming from the minimally coupled quintic galileon would involve at least three gravitons.}. One can show however, that the corresponding term can always be put into the form in which there are no more than four scalars with a single derivative acting on them. Indeed, using the expression for $G_{\mu\nu}$ in terms of the antisymmetric symbol, we have
\beq
(\p\phi)^4 G_{\mu\nu}\nabla^\mu\nabla^\nu\phi \sim \frac{1}{4}(\p\phi)^4 \epsilon_{\mu\alpha\rho\lambda} \epsilon_{\nu\beta\sigma\lambda}\p_\alpha\p_\beta h_{\rho\sigma} (2\p_\mu h_{\gamma\nu}-\p_\gamma h_{\mu\nu})\p_\gamma\phi~.
\eeq
This form makes it straightforward to convince oneself, that the factors involving two gravitons in the above expression collect into either a total derivative, or a total derivative up to corrections involving at most four factors of $\p\phi$:
\be
\begin{split}
\epsilon_{\mu\alpha\rho\lambda} \epsilon_{\nu\beta\sigma\lambda}\p_\alpha\p_\beta h_{\rho\sigma} \p_\mu h_{\gamma\nu}&=\p_\mu\(\epsilon_{\mu\alpha\rho\lambda} \epsilon_{\nu\beta\sigma\lambda}\p_\alpha\p_\beta h_{\rho\sigma} h_{\gamma\nu}\) \\
(\p\phi)^4 \p_\gamma\phi~\epsilon_{\mu\alpha\rho\lambda} \epsilon_{\nu\beta\sigma\lambda}\p_\alpha\p_\beta h_{\rho\sigma}\p_\gamma h_{\mu\nu}&\sim \frac{1}{2} (\p\phi)^4 \p_\gamma\phi~\p_\gamma \(\epsilon_{\mu\alpha\rho\lambda} \epsilon_{\nu\beta\sigma\lambda}\p_\alpha\p_\beta h_{\rho\sigma} h_{\mu\nu}\)~.
\end{split}
\ee
At this point, all that remains is to do a single partial integration to complete the proof of the statement, made in the beginning of the present section.

\section{Spectrum of (near-) de Sitter space}
\label{appB}
Exact de Sitter space saturates the Null Energy Condition (NEC), which seems to be one of the most robust ways of covariantly expressing the requirement of  energy positivity in general relativity\footnote{This is unlike e.g. the Strong Energy Condition (SEC), which is known to be violated both in the early and in the late universe.}. This seems to be supported by the fact, that at least in the simplest effective theories --- those that obey the standard derivative expansion with non-degenerate leading-order effects --- NEC-violation inevitably implies either ghost or gradient instability at short distances \cite{Dubovsky:2005xd}. Moreover, the exact dS limit of the corresponding backgrounds can only be stable over a finite time, necessarily suffering from Jeans-like instabilities \cite{Creminelli:2006xe}. Furthermore, there is additional, purely quantum evidence against the existence of exact de Sitter spacetime \cite{Dvali:2014gua}. In view of these facts, it is important to understand the most general conditions for stability of dS as well as its small deformations.

In this appendix, we study this question for the effective theory governed by WBG invariance. Since these theories lead to second-order equations of motion, the dispersion relation for the scalar perturbations will have a $\omega^2 \propto \vec k^2$ form.
Moreover, as noted above, the operators $\bar M^2$ and $\tilde m^2$ are redundant, and can be removed by a disformal transformation on the tensor mode; one can therefore concentrate on the only remaining quadratic operators: $\delta N^2$ and $\delta N\delta K$ in the effective theory \eqref{s_pi}.
It is convenient to fix the unbroken spatial diffs so as to put the $3D$ metric in the following form \cite{Maldacena:2002vr}
\beq
g_{ij}=a^2(t)[(1+2\zeta)\delta_{ij}+\gamma_{ij}],\qquad \p_i\gamma_{ij}=\gamma_{ii}=0~,
\eeq
where $\zeta$ and $\gamma$ capture the physical scalar and tensor perturbations. Modes of different helicity do not mix on the homogeneous and isotropic backgrounds considered here, so that we can discard the helicity-1 part of the shift altogether, setting $N^i\equiv \delta^{ij}\p_j\beta$. 
In the effective theory for perturbations \eqref{s_pi}, both the lapse and the shift can be integrated out from their respective equations of motion\footnote{For the purposes of deriving the quadratic action for $\zeta$ only the linearized version of these equations matters. This is because  the contribution of higher-order terms in $\delta N$ and $\beta$ to the free  action for the curvature perturbation vanishes on-shell.}:
\begin{align}
\label{lapseeq}
\mpl^2 \bigg[{}^{(3)}\!R+K^2-K^{ij}K_{ij}+\frac{2}{N^2}\dot H-2(3H^2+\dot H) \bigg]+2M^4\delta N-2\hatM^3(\delta K - 3H\delta N)&=0~,  \\
\label{shifteq}
\nabla_i \bigg[\mpl^2\(K^i_{~j}-\delta^i_j K\) -\delta^i_j \hatM^3 \delta N-\bar M'^2 \delta K^i_{~j}-\bar M^2 \delta K \delta^i_j  \bigg]&=0~,
\end{align}
where the lapse perturbation is denoted by $\delta N\equiv N-1$ and all EFT coefficients except for $M^4$ and $\hatM^3$ are set to zero for the reasons outlined above.
Solving these equations for $\delta N$ and $\psi$ to linear order and substituting the solutions back into \eqref{s_pi}, one finds the following quadratic action for the curvature perturbation
\beq
S_\zeta = \int d^4 x~a^3 \bigg [ A(t) \dot \zeta^2-B(t)\frac{(\vec \nabla\zeta)^2}{a^2}     \bigg ]~,
\eeq
where
\begin{align}
A &= \frac{\mpl^2 (-4\mpl^4 \dot H+3 \hatM^6+2 \mpl^2 M^4)}{(\hatM^3-2\mpl^2 H)^2}, \\
B&= \frac{\mpl^2 (-4\mpl^4 \dot H+2 \mpl^2 H\hatM^3 - \hatM^6+2 \mpl^2 \p_t \hatM^3)}{(\hatM^3-2\mpl^2 H)^2}~.
\end{align}
The speed of sound of short-wavelength scalar perturbations is given by $c_s^2=B/A$. The last two expressions agree with those of Ref.~\cite{Creminelli:2010ba} up to a change of operator basis. 

\section{Effective parameters for dark energy}
\label{appC}

Deviations from $\Lambda$CDM for WBG invariant theories can be parametrized in terms of the following time-dependent functions \cite{Bellini:2014fua}\footnote{Here the parameter $\alpha_B$ is defined as $-1/2$ the one defined in Ref.~\cite{Bellini:2014fua}.}
\begin{align}
\alpha_K &=  \frac{ 2 \Lambda_2^4X}{M_*^2 H^2}    \bigg[ G_{2X} + 2 X G_{2XX} -  6Z  (G_{3X} +  X G_{3XX})   \nonumber \\
& + 6 Z^2  \left(\frac{G_{4X}}{X} +8  G_{4XX} + 4X G_{4XXX} \right) + 2 Z^3  \left(3 \frac{G_{5X}}{X} + 7 G_{5XX} + 2 X G_{5XXX} \right) \bigg] \; , \label{alphaK} \\
\alpha_B & =  \frac{  \Lambda_2^4 X }{M_*^2 H^2} Z \bigg[  G_{3X} - 4  Z \left( \frac{G_{4X}}{X} + 2  G_{4XX} \right) -  Z^2 \left(3 \frac{G_{5X}}{X} + 2  G_{5XX} \right) \bigg] \;, \label{alphaB} \\
\alpha_M & =  \frac{ 2 \Lambda_2^4 X}{M_*^2 H^2 } Z^2 \bigg[   \frac{\dot H Z}{H^2 }  \frac{G_{5X}}{X}    + \frac{ \ddot \phi  }{\Lambda_3^3}  \left( 2 \frac{G_{4X}}{X} + 4  G_{4XX} + 3 Z \frac{G_{5X}}{X} + 2 Z  G_{5XX} \right) \bigg]  \;, \label{alphaM} \\
\alpha_T & =  \frac{ 2 \Lambda_2^4 X}{M_*^2 H^2} Z^2 \bigg[    2 \frac{G_{4X}}{X} + \bigg( Z  - \frac{  \ddot \phi }{\Lambda_3^3 }  \bigg) \frac{G_{5X}}{X} \bigg]\;. \label{alphaT}
\end{align}

\bibliographystyle{utphys}
\addcontentsline{toc}{section}{References}
\bibliography{eftinf}

\end{document}